\documentclass[fleqn,usenatbib]{mnras}

\usepackage{newtxtext,newtxmath}

\usepackage[T1]{fontenc}
\usepackage{color}
\usepackage{tabularray}
\usepackage{multirow}
\usepackage{colortbl}
\usepackage[thinlines]{easytable}
\usepackage{booktabs}
\usepackage{adjustbox}
\usepackage{array,booktabs}
\usepackage[table]{xcolor}

\definecolor{Bittersweet}{rgb}{1,0.372,0.372}
\definecolor{AzureRadiance}{rgb}{0,0.631,1}
\DeclareRobustCommand{\VAN}[3]{#2}
\let\VANthebibliography\thebibliography
\def\thebibliography{\DeclareRobustCommand{\VAN}[3]{##3}\VANthebibliography}

\usepackage{graphicx}	
\usepackage{amsmath}	

\title[Tully-Fisher relation: scatter \& radial variation]{The Tully-Fisher relation from SDSS-MaNGA: Physical causes of scatter and variation at different radii}

\author[A. Ristea et al.]{A. Ristea,$^{1,2}$\thanks{E-mail: andrei.ristea@icrar.org (AR)}
L. Cortese,$^{1,2}$
A. Fraser-McKelvie,$^{1,2,3}$
B. Catinella,$^{1,2}$
J. van de Sande,$^{2,4}$
S. M. Croom,$^{2,4}$
\newauthor
and A. M. Swinbank$^{5}$ 
\\
$^{1}$International Centre for Radio Astronomy Research, The University of Western Australia, 35 Stirling Highway, Crawley WA 6009, Australia\\
$^{2}$ARC Centre of Excellence for All Sky Astrophysics in 3 Dimensions (ASTRO 3D), Australia\\
$^{3}$ European Southern Observatory, Karl-Schwarzschild-Straße 2, 85748 Garching, Germany \\
$^{4}$Sydney Institute for Astronomy (SIfA), School of Physics, The University of Sydney, NSW 2006, Australia \\
$^{5}$ Centre for Extragalactic Astronomy, Department of Physics, Durham University, South Road, Durham, DH1 3LE, UK \\
}

\date{Accepted XXX. Received YYY; in original form ZZZ}

\pubyear{2015}

\begin{document}
\label{firstpage}
\pagerange{\pageref{firstpage}--\pageref{lastpage}}
\maketitle

\begin{abstract}
The stellar mass Tully-Fisher relation (STFR) and its scatter encode valuable information about the processes shaping galaxy evolution across cosmic time. However, we are still missing a proper quantification of the STFR slope and scatter dependence on the baryonic tracer used to quantify rotational velocity, on the velocity measurement radius and on galaxy integrated properties. We present a catalogue of stellar and ionised gas (traced by H$\rm{\alpha}$ emission) kinematic measurements for a sample of galaxies drawn from the MaNGA Galaxy Survey, providing an ideal tool for galaxy formation model calibration and for comparison with high-redshift studies. We compute the STFRs for stellar and gas rotation at 1, 1.3 and 2 effective radii ($R_{\rm{e}}$). The relations for both baryonic components become shallower at 2$R_{\rm{e}}$ compared to 1$R_{\rm{e}}$ and 1.3$R_{\rm{e}}$. We report a steeper STFR for the stars in the inner parts ($\leqslant$1.3$R_{\rm{e}}$) compared to the gas. At 2$R_{\rm{e}}$, the relations for the two components are consistent. 
When accounting for covariances with integrated $v/\sigma$, scatter in the stellar and gas STFRs shows no strong correlation with: optical morphology, star formation rate surface density, tidal interaction strength or gas accretion signatures.
Our results suggest that the STFR scatter is driven by an increase in stellar/gas dispersional support, from either external (mergers) or internal (feedback) processes. No correlation between STFR scatter and environment is found. Nearby Universe galaxies have their stars and gas in statistically different states of dynamical equilibrium in the inner parts ($\leqslant$1.3$R_{\rm{e}}$), while at 2$R_{\rm{e}}$ the two components are dynamically coupled. 


\end{abstract}



\section{Introduction}
The Tully-Fisher relation (TFR, \citealt{tully_new_1977}) was  introduced as an empirical correlation between the \ion{H}{i} velocity widths and absolute magnitudes of spiral galaxies. Following its discovery, the TFR has been used extensively as a secondary distance indicator (e.g. \citealt{aaronson_distance_1986, pierce_th_1988, tully_distances_2000, freedman_final_2001}), as a means of measuring the peculiar distance field for cosmological applications \citep{willick_maximum_1997, courteau_shellflow_2000, boubel_preprint_2023, tully_cosmicflows-4_2023}, or for inferring information about the relative contribution of dark-to-luminous matter on galaxy kinematics (e.g. \citealt{persic_dark_1988, gnedin_dark_2007, dutton_revised_2007}).
Early observational studies of the TFR have found a dependence of this relation on the wavelength band in which the magnitude is measured \citep{tully_global_1998, haynes_i-band_1999, verheijen_ursa_2001, courteau_scaling_2007, masters_2mtf_2008}, hinting that a more fundamental correlation can be found by considering an integrated quantity such as total stellar or baryonic mass \citep{mcgaugh_baryonic_2000}. The stellar mass TFR (STFR), together with the mass-size relation, constitute a fundamental set of correlations for galaxies with a global rotational component, akin to the fundamental plane of elliptical galaxies.

Studies of the cosmological origin of the STFR using simulation-based work \citep{steinmetz_cosmological_1999, koda_origin_2008, desmond_tullyfisher_2015} have concluded that the relation can be explained by hierarchical models, although the importance of self-regulating mechanisms such as supernovae feedback and star formation have been found to be critical ingredients in reproducing the STFR \citep{mo_formation_1998, elizondo_self-regulating_1999, somerville_semi-analytic_1999,  buchalter_galactosynthesis_2001, lagos_effects_2008, torrey_model_2014}. Given its fundamental character for galaxy assembly, the STFR represents one of the main calibrations that simulations must match. \cite{ferrero_size_2017} used the Evolution and Assembly of GaLaxies and their Environments (\texttt{EAGLE}, \citealt{schaye_eagle_2015, crain_eagle_2015}) suite of cosmological hydrodynamical simulations to reproduce the STFR, finding a good agreement with observations when using models that match both the abundance and size of galaxies as a function of stellar mass. 

However, the comparison of STFRs produced by simulations with observational-based works (e.g. \citealt{goddy_comparison_2023}) is impeded by the lack of a large representative sample of galaxies with a variety of optical/kinematic morphologies and star forming properties, and spanning various environments, to compare with. Previous STFR studies including early-type (\citealt{heijer_h_2015}) or S0 (\citealt{neistein_tully-fisher_1999}) galaxies, or indeed a variety of optical morphologies  (\citealt{cortese_sami_2014}) typically suffer from low number statistics (<100 objects) in their early-type populations. 

Recent observational studies of the STFR have typically focused on finding the most fundamental correlation that minimises the scatter (e.g. \citealt{courteau_optical_1997, stone_intrinsic_2021, arora_manga_2023}), and in doing so have preferentially selected rotationally supported late-type galaxies.
It is a well-established fact that the baryonic mass TFR is characterised by very low (< 0.08 dex) scatter \citep{mcgaugh_baryonic_2000, trachternach_baryonic_2009, hall_investigation_2012, glowacki_baryonic_2020, mcquinn_turndown_2022}, and its residuals show no correlations with galaxy structural parameters (\citealt{lelli_small_2016}). It has been argued by \cite{lelli_baryonic_2019} that the baryonic mass TFR represents the most fundamental scaling relation of galaxy discs, more fundamental than the angular momentum-galaxy mass relation. 

Furthermore, studies of the TFR have found disagreement between results from rotational velocities at different radii. Typically, the minimum scatter in velocity at fixed stellar/baryonic mass was found when using galaxy samples with flat rotation curves (RCs) in the outer edges, and employing the velocity of the flat part (e.g. \citealt{lelli_baryonic_2019}). The radial dependence of the STFR has been analysed extensively by \cite{yegorova_radial_2007}, who reported a suite of different relations at various scale radii, with slope increasing and scatter decreasing as the probed radius increases (albeit with a relation cast in terms of magnitude, only selecting Sb-Sd galaxies and reaching up to 1.2 effective radii). This set of different relations is a result of rotation curve (RC) shape changing with stellar mass (\citealt{catinella_template_2006}). Furthermore, understanding the radial dependence of the STFR is of particular importance for comparison with high-redshift studies, where a diverse range of radii are used for velocity measurement, probing different scales of the light distribution (e.g. \citealt{tiley_kmos_2016}, \citealt{ubler_evolution_2017} ).

The larger scatter in the local STFR compared to the baryonic relation has been theorised to encode information about the effect of galaxy evolutionary processes such star formation in the disc (see e.g. \citealt{buchalter_galactosynthesis_2001}). Focusing on the scatter in the STFR, \cite{kannappan_physical_2002} have only found shallow correlations with $B-R$ color index and H$\rm{\alpha}$ emission line width. Similar results have been reported by \cite{pizagno_tully-fisher_2007}, who theorised that the scatter is driven by the ratio of dark-to-luminous matter. 
Other shallow correlations of STFR residuals have been reported with disc sizes \citep{reyes_calibrated_2011} and environment \citep{ouellette_spectroscopy_2017}. Analysing 16 elliptical galaxies from the $\rm{ATLAS^{3D}}$ project \citep{cappellari_atlas3d_2011}, \cite{heijer_h_2015} reported that the residuals of this sample with respect to the STFR computed for spirals correlate with mass-to-light ratio, suggesting that the offset is driven by different stellar populations. The effect of galactic bars on kinematics has been considered by \cite{courteau_tully-fisher_2003}, who concluded that these morphological features have no effect on the location of galaxies with respect to the STFR. More recently, \cite{bloom_sami_2017} studied the STFR using the Sydney-AAO Multi-object Integral-field spectrograph (SAMI) Galaxy Survey \citep{croom_sydney-aao_2012, bryant_sami_2015}, reporting a correlation between the vertical scatter and H$\rm{\alpha}$ asymmetry as well as a preference for galaxies with photometric-to-kinematic angle misalignments to be found below the STFR. It is however uncertain what physical processes cause the gas asymmetries in objects scattered below the relation.

It was previously found that the scatter in the STFR can be significantly reduced by applying a correction based on a combination of the velocity dispersion and the rotational velocity component of the respective baryonic tracer. Using a sample of 544 galaxies with strong emission lines, \cite{kassin_stellar_2007} have reported a tight relation between stellar mass and the $S_{0.5}=\sqrt{0.5v_{\rm{rot}}^2 + \sigma^2}$ parameter relating rotational velocity ($v_{\rm{rot}}$) and velocity dispersion ($\sigma$), with little evolution over the redshift range $0.1 < z < 1.2$. The same result has been recovered in the nearby Universe by \cite{cortese_sami_2014} using data for 244
galaxies from the SAMI Galaxy Survey, who  reported that all galaxies regardless of morphology can be placed on a relation between stellar mass and  $S_{0.5}$. Importantly, \cite{cortese_sami_2014} found this to be the case for both the stellar and ionised gas (H$\rm{\alpha}$) kinematics, suggesting that no sample pruning is necessary. 

The results of \cite{cortese_sami_2014} complemented the findings of \cite{catinella_galex_2012} who found that discs and spheroids can be brought to the same dynamical relation by applying a correction to the velocity dispersion that depends on a galaxy's concentration index. 
Similar results have also been recovered by semi-analytical models of galaxy formation implemented by \cite{tonini_fundamental_2014}, who concluded that the ratio of dispersional to rotational motion $\sigma/v$ is a good tracer of the hierarchical assembly history of galaxies and forms a fundamental plane together with galaxy luminosity and rotational velocity.

These studies have demonstrated that the scatter in the TFR at fixed stellar/baryonic mass can be reduced by replacing the \textit{rotational velocity} of a given baryonic tracer with its \textit{circular velocity}. This kinematic measure represents an estimate of the maximum velocity that baryonic particles can have at a given radius if their entire energy budget is used for ordered rotation, and is estimated observationally by taking into account the velocity dispersion component corresponding to non-circular motions (e.g. \citealt{ubler_evolution_2017}). The difference between using rotational versus circular velocities in STFR computation is expected to make a larger difference for stellar kinematics, with stars being able to form dispersion-supported systems given their collisionless nature, and for high-redshift studies tracing H$\rm{\alpha}$, where gas turbulence and the presence of random motions is significant  (e.g. \citealt{tiley_kmos_2016} , \citealt{ubler_evolution_2017}). 

However, it remains unclear whether the scatter in the STFR is entirely specified by considering the velocity dispersion component of a respective baryonic tracer, or whether the distribution of stellar mass also plays an independent role. Furthermore, the role of environment, tidal interactions, morphological features (bars/rings), star formation or gas accretion in driving the scatter in the STFR is still poorly understood.

In this work, we compute the STFR at different multiples (1, 1.3 and 2) of the effective radius $R_{\rm{e}}$ for a representative sample of galaxies drawn from the final data release of the Sloan Digital Sky Survey IV - Mapping Nearby Galaxies at Apache Point Observatory (SDSS IV - MaNGA, \citealt{bundy_overview_2015, drory_manga_2015}) Galaxy Survey Data Release 17 (DR17, \citealt{ abdurrouf_seventeenth_2022}). We make measurements at 1.3$R_{\rm{e}}$ as this is the radius where the rotational velocity of a purely exponential disc galaxy reaches its maximum, while 1$R_{\rm{e}}$ and 2$R_{\rm{e}}$ are probing the inner (potentially bulge-dominated) and outer (flat, rising or declining, depending on the galaxy's stellar mass and morphology; see \citealt{yoon_rotation_2021}) part of the velocity profile. Our study provides a benchmark for comparison with cosmological simulations of galaxy formation and with high-redshift studies of the STFR. We undertake a comparative analysis of the STFR for stellar and ionised gas rotation (traced by H$\rm{\alpha}$ emission), and analyse the correlation between STFR residuals and galaxy optical/kinematic morphology, star formation rate surface density, environment, signatures of gas accretion and presence of bars/rings.

This paper is structured as follows: Section 2 describes the galaxy sample used and kinematic extraction method; Section 3 presents the STFR for ionised gas and stellar rotation at different radii, and places our findings in the context of galaxy evolution; Section 4 analyses the physical causes of scatter in the STFR for stellar and ionised gas rotation individually and discusses the implication of these results for our understanding of the processes that shape the evolution of these baryonic components; Section 5 summarises our findings and provides concluding remarks. Throughout this paper, we assume a flat $\rm{\Lambda}$CDM concordance cosmology: $H_0=70\ \rm{km\ s^{-1}\ Mpc^{-1}}$, $\Omega_0=0.3$, $\Omega_{\Lambda}=0.7$.

\section{Sample selection and Kinematic extraction}
\label{sec:data_and_methods}

\subsection{The MaNGA galaxy survey}
\label{sec:MaNGA_survey}
The MaNGA Galaxy Survey \citep{bundy_overview_2015, drory_manga_2015} is an SDSS-IV Project \citep{blanton_sloan_2017} employing the Baryon Oscillation Spectroscopic Survey (BOSS) spectrographs \citep{smee_multi-object_2013} on the 2.5m telescope at Apache Point Observatory \citep{gunn_25_2006}. The $17^{\rm{th}}$ and final data release of MaNGA (MaNGA DR17,  \citealt{abdurrouf_seventeenth_2022}) contains integral field spectroscopic observations of 10,010 unique galaxies in the redshift range $0.01 < z < 0.15$, with a roughly uniform distribution in $\log(M_{\star})$ between $5\times 10^{8} \leqslant M_{\star} \leqslant 3 \times 10^{11} \rm{M_{\odot}}$ $h^{-2}$ (\citealt{wake_sdss-iv_2017}), reduced by the MaNGA data reduction pipeline (DRP, \citealt{law_observing_2015}). 

The MaNGA Primary+ sample \citep{yan_sdss-ivmanga_2015, wake_sdss-iv_2017} contains spatial coverage out to $\sim\ 1.5$ effective radii ($R_{\rm{e}}$) for $\sim\ 66$ per cent of the entire DR17 Sample. The remaining galaxies constitute the secondary sample, for which observations reach $\sim\ 2.5\ R_{\rm{e}}$. MaNGA DR17 includes the release of derived spectroscopic products (stellar kinematics and emission-line diagnostic maps) from the MaNGA data analysis pipeline (DAP, \citealt{belfiore_sdss_2018, westfall_data_2019}), provided as a single data cube per galaxy \citep{yan_sdss-iv_2016}. In this work, we analyse the 2D kinematic maps produced by the MaNGA DAP: stellar and ionised gas (traced by H$\rm{\alpha}$ emission) rotational velocities (and their associated error maps), and velocity dispersions.

\subsection{Galaxy properties}
\label{sec:gal_prop}
We make use of several galaxy physical and environmental properties to compute the STFR and study the physical causes of its scatter: stellar mass ($\boldsymbol{{M_{\star}}}$), star formation rate (\textbf{SFR}), r-band integrated S\'ersic index ($\boldsymbol{n_{\rm{s}}}$), elliptical Petrosian 50 per cent light radius in the r-band ($\boldsymbol{R_{\rm{e}}}$), semi-minor to semi-major axis ratio ($\boldsymbol{b/a}$), stellar and ionised gas velocity to dispersion ratio \textbf{$\boldsymbol{(v/\sigma)_{\rm{N}R_{\rm{e}}}}$} (integrated within N$\times R_{\rm{e}}$, N = 1, 1.3, 2), \textbf{T-type}, galaxy group membership (\textbf{isolated, satellite, central}), group tidal strength parameter \textbf{$\boldsymbol{Q_{\rm{group}}}$}, galaxy \textbf{morphological feature indicator} (i.e the presence of a bar or ring). The sources and/or computation methods of these parameters are as follows:

\begin{itemize}
\item $\boldsymbol{{M_{\star}}}$ and \textbf{SFR}: Extracted from the \textit{GALEX}-Sloan-\textit{WISE} Legacy Catalogue 2 (GSWLC-2, \citealt{salim_lessigreatergalexlessigreater_2016, salim_dust_2018}). This catalogue employs the deepest available \textit{GALEX} photometry to compute \textbf{SFR} and $\boldsymbol{M_{\star}}$ using the spectral energy distribution (SED)-fitting Code Investigating GALaxy Evolution (CIGALE; \citealt{burgarella_star_2005}; \citealt{noll_analysis_2009}; \citealt{boquien_cigale_2019}), and employing a Chabrier initial mass function (\citealt{chabrier_galactic_2003}). This catalogue was matched to the MaNGA DR17 sample using a sky match with a maximum separation of $2^{\prime\prime}$. Out of the galaxies in MaNGA DR17, 8637 have \textbf{SFR} and $\boldsymbol{M_{\star}}$ values in GSWLC-2.

\item $\boldsymbol{R_{\rm{e}}}$, $\boldsymbol{n_{\rm{s}}}$ and $\boldsymbol{b/a}$: Extracted from MaNGA's DRP summary table \texttt{drpall$\_$v3$\_$1$\_$1}. These parameters are compiled from the NASA Sloan Atlas catalogue \citep{blanton_improved_2011}. 
In this work, we use half-light radii computed from r-band SDSS imaging using the elliptical Petrosian method (with the seeing point-spread function (PSF) accounted for), as opposed to results from a S\'ersic fit. The latter method has been shown to suffer more catastrophic failures while also producing $R_{\rm{e}}$ values that are systematically overestimated for galaxies with high concentrations, compared to the elliptical Petrosian method (\citealt{wake_sdss-iv_2017}).   

\item  \textbf{$\boldsymbol{(v/\sigma)_{\rm{N}R_{\rm{e}}}}$} (N = 1, 1.3, 2): Computed as described in equations (1) and (2) from \cite{fraser-mckelvie_beyond_2022}, using the definition of \cite{emsellem_sauron_2007, emsellem_atlas3d_2011} and with the weighted averaging performed over all spaxels within 1$R_{\rm{e}}$, 1.3$R_{\rm{e}}$ and 2$R_{\rm{e}}$ that pass the quality cuts outlined in \cite{fraser-mckelvie_beyond_2022}. 
While this definition of $v/\sigma$ is rarely used in ionised gas kinematic studies, we adopt it here to provide a 1-to-1 match with what is performed for stellar kinematics. Throughout this paper, we largely make use of non-corrected (for beam-smearing and inclination) values of $(v/\sigma)$ for both stars and gas, in order to preserve a consistency between the two tracers, since reliable corrections are only available for the stellar kinematics. In cases where beam-smearing- and inclination- corrected values of $(v/\sigma)$ for stars are used (as specified), we make use of the correction prescriptions of \cite{harborne_recovering_2020} and \cite{emsellem_atlas3d_2011}, respectively.

\item \textbf{T-type}: Extracted from the MaNGA Morphology Deep-Learning DR17 value added catalogue (\citealt{fischer_sdss-iv_2019, dominguezsanchez_sdss-iv_2021}). Similarly to \cite{yoon_rotation_2021}, we use 14 T-type values from -3 to 10: E (-3), E/S0 (-2), S0 (-1), S0/a (0), Sa, Sab, Sb, Sbc, Sc, Scd, Sd, Sdm (1-8), Sm (9), Irr (10). 

\item \textbf{group membership} and $\boldsymbol{Q_{\rm{group}}}$: Extracted from the \textit{GEMA-VAC: Galaxy Environment for MaNGA} Value Added Catalogue. This catalogue contains several environmental quantifications for MaNGA galaxies, as described in \cite{argudo-fernandez_catalogues_2015}, \cite{etherington_measuring_2015} and \cite{wang_elucid_2016}. In this work, we use the metrics separating satellite galaxies from centrals/isolated (identified as either the brightest or most massive galaxy in a group of size $\geqslant$ 2). We also employ the group tidal strength parameter $Q_{\rm{group}}$ which is a measure of the gravitational tidal force that a galaxy feels from its group neighbours and is computed as described in \cite{argudo-fernandez_catalogues_2015}.

\item \textbf{morphological features}: Extracted from the \textit{Galaxy Zoo classifications for MaNGA DR17 galaxies} value added catalogue (see \citealt{willett_galaxy_2013}, \citealt{walmsley_galaxy_2022}). This catalogue provides a probability that a galaxy contains a particular feature based on user identification, and weighting scorers on their accuracy. In this work, we make use of the debiased probability that a given galaxy contains a bar (\texttt{T03$\_$BAR$\_$A06$\_$BAR$\_$DEBIASED}) or ring (\texttt{T08$\_$ODD$\_$FEATURE$\_$A19$\_$RING$\_$DEBIASED}). These values take into account the redshift difference between the galaxies in the classified sample and the fact that higher redshift galaxies are less likely to have a certain feature identified (see \citealt{walmsley_galaxy_2022}  for a full description). We employ the same selection criterion as \cite{fraser-mckelvie_sdss-iv_2020} for labelling a galaxy as having a bar or ring, i.e. if the respective debiased probability is $>$ 0.5. All galaxies with b/a $<$ 0.25 are considered too close to edge-on to have an accurate classification, being labelled as 'ambiguous' for this categorisation and thus not considered in any analysis when a separation based on morphological features is required.

\end{itemize}
We also compute stellar mass and SFR surface densities: $\Sigma_{\star} =  M_{\star}/2\pi R_{\rm{e}}^{2}$ and $\Sigma_{\rm{SFR}} = \rm{SFR}$$/2\pi R_{\rm{e}}^{2}$, respectively. 

We note that the effect of beam-smearing will result in our measured rotational velocities being lower than the real values. The size of this reduction has been previously shown to depend on the ratio of a galaxy's effective radius to the seeing full width at half maximum (FWHM) of the  PSF, $R_{\rm{e}}$/FWHM (see e.g. \citealt{johnson_kmos_2018}). To mitigate this effect while also not biasing our final sample towards large $R_{\rm{e}}$ values, we only consider galaxies with $R_{\rm{e}}$ / FWHM $\geqslant$ 1 in any further analysis (7916 objects, 91 per cent of out initial sample). The average FWHM for the MaNGA DR17 sample is $2.5^{\prime \prime}$.

\begin{figure*}
	\centering
	\includegraphics[width=\linewidth]{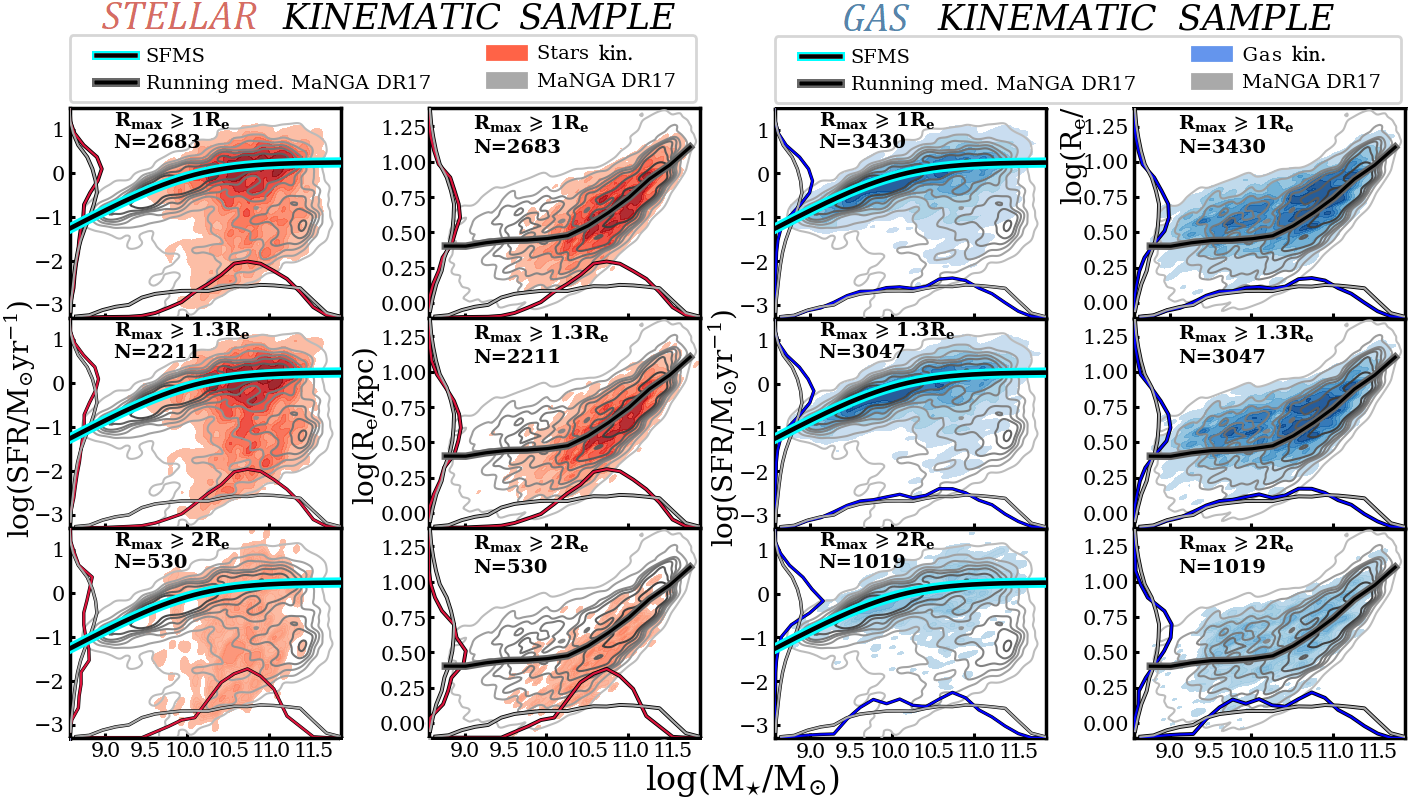}
    \caption{A description of our \textbf{\textcolor{red}{stellar}} (\textbf{left columns}) and \textbf{\textcolor{blue}{gas}} (\textbf{right columns}) kinematic galaxy samples used in this work, in terms of the SFR-$M_{\star}$ and $R_{\rm{e}}-M_{\star}$ planes. Contours show the number density of galaxies. The rows show the sub-samples with kinematics that reach (from top to bottom) at least 1$R_{\rm{e}}$, 1.3$R_{\rm{e}}$ and 2$R_{\rm{e}}$, as indicated on each panel. The number of galaxies in each sub-sample is shown on each panel. Each sub-sample is shown in comparison with the full MaNGA DR17 parent sample, presented as grey contours. The black-cyan line in the SFR-$M_{\star}$ plots shows the star forming main sequence (SFMS, as computed by \citealt{fraser-mckelvie_sami_2021}). The black line on the $R_{\rm{e}}-M_{\star}$ plots shows the running median of the MaNGA DR17 parent sample. The insert lines on the sides of each plot show the normalised probability distribution function of the respective parameter (SFR, $M_{\star}$ or $R_{\rm{e}}$), for the sample with matching color.} 
    \label{fig:sample_plot}
\end{figure*}

\subsection{Quality cuts and kinematic position angles}
\label{sec:Qc_Pa}

In this work, we extract the rotational velocities of stars and ionised gas at different radii (1$R_{\rm{e}}$, 1.3$R_{\rm{e}}$ and 2$R_{\rm{e}}$) from resolved 2D kinematic maps. In addition to the MaNGA spaxel masks, we exclude any spaxels in the stellar and gas rotational velocity maps ($v_{\rm{rot}}$) where:
\begin{itemize}
    \item $\rm{(S/N) < 5}$ ;
    \item error$(v_{\rm{rot}}) > 0.5\times|v_{\rm{rot}}|+15\ \rm{km\ s^{-1}}$. 
\end{itemize}
The S/N ratios for spaxels in the ionised gas maps are computed as the ratio between the H$\rm{\alpha}$ line intensity and its associated uncertainty.
The above cuts are applied \textit{individually} to the stellar and gas velocity maps and ensure that our analysis does not include spaxels with high random uncertainties that might bias our kinematic measurements.

Following the application of quality cuts, we compute stellar and gas kinematic position angles (PAs) using the method described in \cite{ristea_sami_2022}. The PA calculation is done for all the galaxies that have $M_{\star}$ and SFR measurements, $R_{\rm{e}}/\rm{FWHM}$ $\geqslant$ 1 (Section \ref{sec:gal_prop}), and $\rm{N_{spaxels}}$ $\geqslant$ 50 left in \textit{either} the stellar or gas velocity maps after the application of quality cuts. Briefly, we compute kinematic PAs using both the Radon transform \citep{stark_sdss-iv_2018} and Kinemetry \citep{krajnovic_kinemetry_2006} methods. The Radon transforms method is non-parametric and computes kinematic PAs by integrating the difference between the mean and individual spaxel velocities along a given line passing through the kinematic centre. The kinematic PA is then given by the line for which this integral is minimised. The Kinemetry method is implemented by the \texttt{FIT$\_$KINEMATIC$\_$PA}\footnote{\url{https://www-astro.physics.ox.ac.uk/~mxc/software/}} routine described by \cite{cappellari_sauron_2007} and \cite{krajnovic_atlas3d_2011}. This method assumes that the velocity profile is bi-antisymmetric with respect to the minor and major kinematic axes. Velocity maps are rotated in $1^{\rm{o}}$ increments and at each step a bi-antisymmetric model map is computed, which is then subtracted from the rotation map. The difference between the data and the bi-antisymmetric model is minimised when the kinematic PA is aligned with the horizontal axis.

Stellar and gas kinematic PAs are computed for 7888 and 6290 galaxies, respectively. We calculate the difference between the PAs produced by the two methods, i.e. $\Delta \rm{PA}_{\rm{Kinemetry-Radon}}$, for stars and gas \textit{individually}. The $16^{\rm{th}}$ and $84^{\rm{th}}$ percentiles of $\Delta \rm{PA}_{\rm{Kinemetry-Radon}}$ are $(-12^{\rm{o}},11^{\rm{o}})_{\rm{stars}}$ and $(-11^{\rm{o}},9^{\rm{o}})_{\rm{gas}}$ for stars and gas, respectively. We visually inspect all stellar and gas velocity maps for which $\Delta \rm{PA}_{\rm{Kinemetry-Radon}}$ is outside these values and identify the PA computation method which best follows the direction of the maximum velocity gradient. We also identify all  cases with no global rotation pattern or highly disturbed kinematics (indicative of ongoing mergers or interactions). These objects (1267 for stellar and 888 for gas kinematics) are excluded from any further analysis, leaving 6621 and 5402 galaxies with stellar and gas kinematic PAs, respectively. Finally, from our stellar ($\rm{PA}_{\rm{st}}$) and gas ($\rm{PA}_{\rm{g}}$) kinematic position angle measurements, we calculate the misalignments between the kinematic axes of the two baryonic components as $\Delta \rm{PA}_{\rm{st-g}} = |\rm{PA}_{\rm{st}} - \rm{PA}_{\rm{g}}|$.

\subsection{Rotation curve extraction and centering}
\label{sec:rc_extract}

We extract stellar and gas RCs along the major kinematic axis for the galaxies with kinematic PA measurements (Section \ref{sec:Qc_Pa}). We rotate velocity maps so that the major kinematic axes are aligned with the x-axis and select a slit of width equal to the median FWHM for our sample ($2^{\prime\prime}$), corresponding to 4 pixels. The same procedure is applied to velocity error maps to extract uncertainties in rotational velocities.

To account for shifts of the kinematic centre along the major axis ($r_{\rm{off}}$), we fit the extracted stellar and gas RCs with the velocity model presented in \cite{yoon_rotation_2021}. We further include a parameter to account for global offsets in the rotation curve in velocity, $v_{\rm{off}}$, corresponding to the velocity at a radius equal to 0:

\begin{equation}
    V(r) - v_{\rm{off}} = V_{\rm{c}}\ \tanh  \Bigl(  \frac{r-r_{\rm{off}}}{R_{\rm{t}}}  \Bigl) \ + \ s_{\rm{out}}(r-r_{\rm{off}}),
	\label{eq:yoon_fit}
\end{equation}

\noindent where $V_{\rm{c}}$, ${R_{\rm{t}}}$ and  $s_{\rm{out}}$ represent the coefficient of the $\tanh$ term equal to the maximum velocity at $s_{\rm{out}}=0$, the turnover radius where the $\tanh$ term begins to flatten, and the slope of the RC at large radii $r\gg R_{\rm{t}}$, respectively. The above model is fitted with $V_{\rm{c}}$, ${R_{\rm{t}}}$, $s_{\rm{out}}$, $r_{\rm{off}}$ and $v_{\rm{off}}$ as free parameters.

In this study, we measure rotational velocities at fixed scale radii, i.e 1$R_{\rm{e}}$, 1.3$R_{\rm{e}}$ and 2$R_{\rm{e}}$, which require a knowledge of the kinematic centre. Given this necessity, it is advisable to exclude rotational profiles with highly uncertain kinematic centres, either due to physical reasons that disturb the kinematics, or due to measurement errors or poor data quality. As such, we exclude the stellar and/or gas kinematics of galaxies where the offset in the respective rotation curve is $r_{\rm{off}} > 0.3 R_{\rm{e}}$. This value is a conservative cut which excludes the galaxies outside the $2^{\rm{nd}}$ and $98^{\rm{th}}$ percentiles (corresponding to $\approx2\sigma$ for a normal distribution) of the $r_{\rm{off}}/R_{\rm{e}}$ distributions for both stellar and gas rotation, equal to $(-0.34,0.35)_{\rm{stars}}$ and $(-0.32, 0.32)_{\rm{gas}}$ for the two components. This cut corresponds to an exclusion of galaxies with $|r_{\rm{off}}|\  \gtrsim 1.36^{\prime\prime}$ for stellar kinematics, and $|r_{\rm{off}}|\  \gtrsim 1.17^{\prime\prime}$ for gas. Following this selection, we centre our rotation curve by subtracting the best-fit $v_{\rm{off}}$ and $r_{\rm{off}}$ values from each RC's velocity and radius.

We correct for the effect of the inclination angle $\theta_{\rm{inc}}$ of each galaxy with respect to the line of sight by dividing our velocity measurements by sin($\theta_{\rm{inc}}$). The inclination angle is calculated as:

\begin{equation}
    {\cos}^2(\theta_{\rm{inc}})=\frac{(b/a)^2 - q^2_0}{1-q^2_0},
	\label{eq:incl_corr}
\end{equation}

where $q_0$ is a galaxy's intrinsic axis ratio. We calculate $q_0$ using the prescription of \cite{bottinelli_h_1983}, which takes into account each galaxy's light distribution parameterised by the T-type. For the galaxies used in this work, the $q_0$ value is between 0.13 and 0.55, with a median of 0.24. The inclination corrections computed using this method are consistent with values obtained when considering $q_{0}=0.2$ for discs (T-type > -1.5) and $q_{0}=0.6$ for ellipticals (T-type $\leqslant$ -1.5). The ratio of inclination correction coefficients obtained using the two $q_0$ prescriptions for the galaxy sample in this work has a median of 0.99 and a standard deviation of 0.07.
The inclination-corrected and centred RCs are then folded in radial space, by taking the average velocity at each radius about the kinematic centre.

From our centred, folded and inclination-corrected stellar and gas RCs, we extract rotational velocities at 1$R_{\rm{e}}$, 1.3$R_{\rm{e}}$ and 2$R_{\rm{e}}$ by taking the average between the velocity measurements at the two closest points on each side of the respective radius. This kinematic extraction approach makes no assumption about the underlying kinematic structure of stars and gas beyond that of rotation in a plane that can be described by the specification of a major kinematic axis. This method also provides consistency with the integrated $v/\sigma$ measurements, which are extracted directly from the kinematics without any prior modelling. As such, this approach is preferred in this work to more parametric methods involving fitting of the 1D or 2D galaxy kinematics. We also tested the approach of re-fitting the RCs with the \cite{yoon_rotation_2021} model of equation \ref{eq:yoon_fit} and extracting the rotational velocities from the best-fit models. The differences between these approaches on our final results are very small, with average ratios (velocity from fit to directly extracted velocity) of $(1.00, 1.00, 1.01)_{\rm{stars}}$ and $(1.01, 1.01, 1.02)_{\rm{gas}}$ between the velocities produced by the two methods at 1$R_{\rm{e}}$, 1.3$R_{\rm{e}}$ and 2$R_{\rm{e}}$ for stars and gas, respectively. For the remaining of this paper, we make use of the rotational velocities measured directly from the RC (i.e. without fitting).

Corrections for inclinations become unreliable for the most face-on systems as $\theta_{\rm{inc}}$ approaches $0^{\rm{o}}$. We apply a conservative cut and discard all galaxies where the inclination correction 1/$\rm{sin(\theta_{\rm{inc}})}$ is larger than 1.7 (corresponding to $\theta_{\rm{inc}}\  \lesssim 36^{\rm{o}}$). We use this threshold as the value for which the distribution in error$(v_{\rm{off}})$ for the stellar kinematic sample reaches an average equal to 10 per cent of the average of the distribution in velocity at $1R_{\rm{e}}$ for the respective sample (for the gas kinematic sample, the corresponding value is 3 per cent). Furthermore, the presence of dust lanes can bias measurements of light-weighted parameters (in this case, integrated stellar and gas $v/\sigma$ ratios) for the most edge-on systems. We test this potential effect on the computed $v/\sigma$ values by splitting our stellar and gas kinematic samples reaching different radii in terms of inclination, at $\theta_{\rm{inc}} = 60^{\rm{o}}$. The medians of $v/\sigma$ distributions for galaxies with inclinations above and below $60^{\rm{o}}$ are always consistent within one standard deviation, for both stellar and gas measurements, at all probed radii. The same result is obtained if we instead change the inclination threshold in $1^{\rm{o}}$ increments between $55^{\rm{o}}$ - $65^{\rm{o}}$.

\subsection{Final samples}
\label{sec:final_sample}

In this work, we select two samples of galaxies with reliable stellar and gas rotational kinematic measurements from the parent MaNGA DR17 sample (hereafter, the \textbf{stellar and gas kinematic samples}). We also consider the sample of galaxies which have both stellar and gas kinematics, i.e. the intersection of the stellar and gas kinematic samples, referred to as the \textbf{common kinematic sample}. Some exclusion criteria have been presented in Sections \ref{sec:gal_prop}-\ref{sec:rc_extract}. Below, we present a compilation of our full sample selection cuts. To be included in our final \textbf{stellar/gas kinematic sample}, a galaxy in MaNGA DR17 must have:
\begin{enumerate}
    \item $M_{\star}$, SFR and $R_{\rm{e}}$ measurements available (8616 galaxies);
    \item $R_{\rm{e}}$ / FWHM $\geqslant\ 1$ (7916 galaxies);
    \item $\rm{N_{\rm{spaxels}}} \geqslant 50$ in stellar or gas velocity maps (7888 and 6290 galaxies with stellar and gas kinematics, respectively);
    \item Has not been identified as having unreliable/highly disturbed kinematics by visual inspection, as described in Section \ref{sec:Qc_Pa} (6621 and 5402 galaxies with stellar and gas kinematics, respectively);
    \item $r_{\rm{off}} < 0.3\ R_{\rm{e}}$ (6193 and 5124 galaxies with stellar and gas kinematics, respectively);
    \item sin$(\theta_{\rm{inc}}) < 1.7$ (3978 and 3445 galaxies with stellar and gas kinematics, respectively);
    \item $R_{\rm{max}} \geqslant 1R_{\rm{e}}$ (or 1.3$R_{\rm{e}}$, 2$R_{\rm{e}}$ - numbers of galaxies in each sub-sample are displayed in the legend of Fig. \ref{fig:sample_plot}).
\end{enumerate}

We noted in Section \ref{sec:gal_prop} that beam-smearing will have an effect on our measured rotational velocities. While this effect is expected to be minimal given our cut in $R_{\rm{e}}$ / FWHM $\geqslant\ 1$, we perform a test by computing the beam-smearing corrections for the H$\rm{\alpha}$ rotational velocity of galaxies in our gas kinematic sample at 1.3$R_{\rm{e}}$ and 2$R_{\rm{e}}$ using the prescription of \cite{johnson_kmos_2018}, noting that no such corrections were computed by those authors for velocities at 1$R_{\rm{e}}$. The mean velocity  corrections at 1.3$R_{\rm{e}}$ and 2$R_{\rm{e}}$ are 1.04 and 1.03 respectively (with standard deviations equal to 0.06 and 0.04). Given that these values are well within measurement errors for our velocities, and the fact that no beam-smearing correction for stellar kinematics is available (although the effect of beam-smearing is expected to be similar), for consistency between stellar and gas kinematics we proceed without applying this correction. 

Our stellar and gas kinematic samples are presented in Fig. \ref{fig:sample_plot} on the SFR$-M_{\star}$ and $R_{\rm{e}}-M_{\star}$ planes, also highlighting the number of galaxies in each sample. In the following, we discuss biases introduced by our sample selection procedure, in terms of the galaxy parameters shown in Fig. \ref{fig:sample_plot}.

For our stellar kinematic sample (red), the selection criteria induce a bias against low-$M_{\star}$ ($\lesssim10^{9.5}\rm{M_{\odot}}$) galaxies at all radii, albeit more pronounced at 2$R_{\rm{e}}$. We use a Kolmogorov-Smirnoff (KS) test to compare the log$(M_{\star})$ distributions for our stellar kinematic and MaGNA DR17 samples. Under the null hypothesis that two distributions are drawn from the same parent distribution, the KS test p value returns the probability that the log$(M_{\star})$ distributions for the two samples can be as different as observed. We reject the null hypothesis for p values < 0.05. This is the case  when comparing the log$(M_{\star})$ distributions of the stellar kinematic samples on the left of Fig. \ref{fig:sample_plot} with that of the MaGNA DR17 parent sample, at all radii (p values $\leqslant\ 4.8\times10^{-28}$). At 2$R_{\rm{e}}$, our stellar kinematic sample lacks coverage of both the low ($\lesssim10^{10}\rm{M_{\odot}}$) and high ($\gtrsim10^{11}\rm{M_{\odot}}$) end of the log$(M_{\star})$ range probed by the MaNGA DR17 sample. 
The gas kinematic sample used in this work (blue) probes the MaNGA DR17 log$(M_{\star})$ distribution better than the stellar one for sub-samples reaching all probed radii, with greater coverage at the lower stellar mass end (log$(M_{\star})<9.5$), albeit still being statistically different (p values $\leqslant\ 1.8\times10^{-11}$).

A comparison with the MaNGA DR17 sample in terms of effective radius reveals that the stellar sample is typically biased against the lower end of the $R_{\rm{e}}$ distribution of our parent sample at 1$R_{\rm{e}}$ and 1.3$R_{\rm{e}}$ (p value < 0.05), unsurprising given our selection impact on the $M_{\star}$ distribution of our stellar kinematic sample. At 2$R_{\rm{e}}$, this selection bias is more pronounced (p value = $1.6\times10^{-9}$), although we are preferentially selecting galaxies in the lower end of the MaNGA DR17 $R_{\rm{e}}$ distribution. Similar results are found in the case of the gas kinematic sample, with the exception that the sub-sample reaching 2$R_{\rm{e}}$ is undersampling both the lower and higher end of our parent sample $R_{\rm{e}}$ distribution.

The stellar kinematic sample's distribution of SFRs compares well with that of the parent MaNGA sample. On the other hand, we note that the gas kinematic sample is biased towards main sequence galaxies for all sub-samples in Fig. \ref{fig:sample_plot} (a KS test comparison with the SFR distribution of the MaNGA DR17 sample yields a p value < 0.05 in all cases). This is due to a lack of extended H$\rm{\alpha}$ emission in quenched galaxies, given that this emission line is generally an instantaneous tracer of star-formation (with the exception of emission originating in low-ionisation nuclear emission regions, or Seyfert objects).

One of the goals of this study is to analyse the correlation between the vertical scatter about the STFR at different radii, and various galaxy properties and environmental metrics. To define a relation with respect to which this scatter is to be calculated, we select a sample of galaxies with rotational dominance ($v/\sigma$ > 0.56, as discussed below) in \textit{both} stars and gas (hereafter, the \textbf{rotator} sample). These objects are expected to have the tightest relation between rotational velocity and stellar mass, as scatter in the STFR has been shown to correlate with dispersional support (stars) and turbulence (gas) (e.g. \citealt{cortese_sami_2014}, \citealt{lelli_baryonic_2019}).

We employ the delimitation of \cite{fraser-mckelvie_beyond_2022} at log($v/\sigma$) = -0.125 ($v/\sigma$ = 0.75, corrected for beam-smearing), derived as a separation between dynamically cold discs and intermediate systems. We note that while both of these classes are \textit{fast rotators} according to the definition of \citealt{cappellari_sauron_2007}, in this work we aim to identify the most rotationally dominated objects as our rotator sample, rather than perform a definitive separation between cold discs and systems in other dynamical states. We re-scale the log($v/\sigma$) threshold to make it applicable for non beam-smearing corrected values. This re-scaling implies multiplying by the ratio between $v/\sigma$ not corrected vs corrected for beam-smearing, which we are able to compute for our stellar kinematic sample.  
The median of this ratio is equal to 0.74 for the stellar kinematic sub-samples reaching all probed radii, with standard deviations equal to 0.12, 0.11 and 0.12 at 1$R_{\rm{e}}$, 1.3$R_{\rm{e}}$ and 2$R_{\rm{e}}$, respectively. For consistency between all sub-samples considered in this work, and between different kinematic tracers, we employ the median correction factor of 0.74, which results in a threshold of $v/\sigma$ = 0.56.

This threshold correction factor is however not unique for galaxies in our sample given their different $R_{\rm{e}}/$FWHM ratios. We test the application of a more stringent separation at $v/\sigma$ = 0.65 (computed by adding the standard deviation to the median threshold correction factor). The effect of this selection on the best-fit STFR parameters presented in Section \ref{sec:TF_kinematic_full} and Section \ref{sec:TF_kinematics_matched} are minimal, with differences always being within uncertainties.

The criteria for selecting our rotator sample are as follows:
\begin{enumerate}
    \item Galaxies must be centrals or isolated, in order to ensure the kinematics of such objects are not affected by group physical processes that might act to disturb kinematics;
    \item Galaxies must be in \textbf{both} the stellar and gas kinematic samples at the respective radius (1$R_{\rm{e}}$, 1.3$R_{\rm{e}}$ or 2$R_{\rm{e}}$);
    \item $(v/\sigma)_{\rm{N}R_{\rm{e}}} \rm{(not\ corrected\ for}$ beam-smearing) > 0.56, N = 1, 1.3, 2. For a galaxy to be included in the rotator sample at 2$R_{\rm{e}}$/1.3$R_{\rm{e}}$, it must also pass the $v/\sigma$ threshold for the values within lower radii (i.e 1$R_{\rm{e}}$ and 1.3$R_{\rm{e}}$ for the rotator sample reaching 2$R_{\rm{e}}$, and 1$R_{\rm{e}}$ for the rotator sample reaching 1.3$R_{\rm{e}}$).
\end{enumerate}

\begin{figure}
	\centering
	\includegraphics[width=\columnwidth]{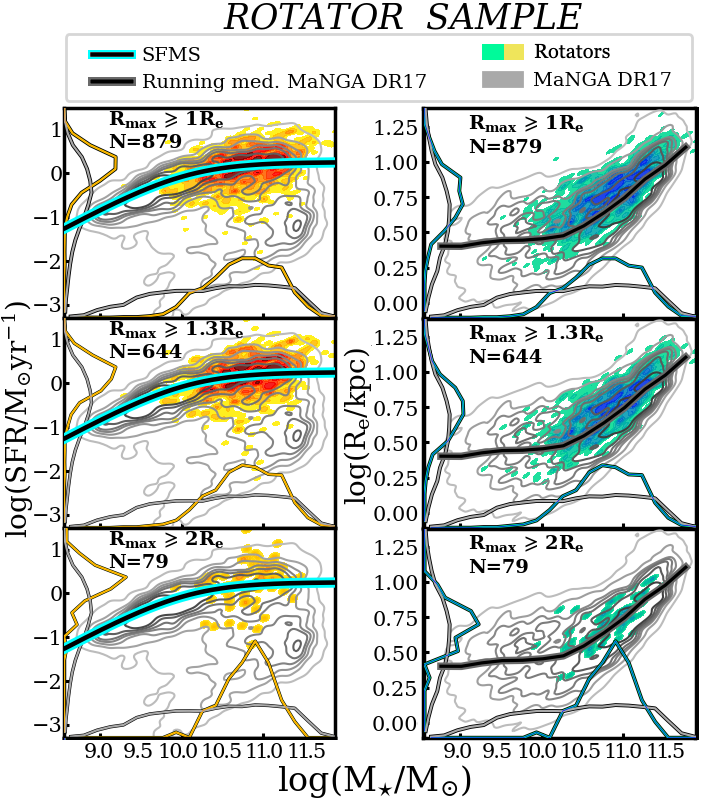}
    \caption{The same as Fig. \ref{fig:sample_plot}, for the rotator sample of galaxies.} 
    \label{fig:sample_plot_rotators}
\end{figure}

The rotator sample is shown on the SFR$-M_{\star}$ and $R_{\rm{e}}-M_{\star}$ planes in Fig. \ref{fig:sample_plot_rotators}. For sub-samples at all radii, these galaxies are typically massive objects (96, 97 and 99 per cent of rotator sample galaxies with kinematics reaching 1$R_{\rm{e}}$, 1.3$R_{\rm{e}}$ and 2$R_{\rm{e}}$ have $M_{\star}>10^{10}\rm{M_{\odot}}$) located around the star forming main sequence (89, 90 and 90 per cent of rotator sample galaxies with kinematics reaching 1$R_{\rm{e}}$, 1.3$R_{\rm{e}}$ and 2$R_{\rm{e}}$ are within 0.5 dex of the star forming main sequence). The size distributions of the selected rotator galaxies are typically sampling the higher end of the MaNGA DR17 sample's distribution (76, 72 and 73 per cent of rotator sample galaxies with kinematics reaching 1$R_{\rm{e}}$, 1.3$R_{\rm{e}}$ and 2$R_{\rm{e}}$ have $R_{\rm{e}}$ larger than 4.14 kpc, corresponding to the MaNGA DR17 median). 

A lack of galaxies with low (<$10^{9.5} \rm{M_{\odot}}$) stellar masses and sizes in the stellar kinematic and rotator samples is due to our need for measuring $v/\sigma$ in order for galaxies to be included in these samples. For our spaxel quality cut of S/N > 5, in the case of stellar kinematics, measuring $v/\sigma$ below $\sim50\ \rm{km\ s^{-1}}$ becomes unreliable (uncertainty > 60 per cent, see Fig. 15 of \citealt{law_observing_2015}). Given the scaling relation between stellar mass and velocity dispersion, this cut will preferentially exclude low-$M_{\star}$ galaxies. The same velocity dispersion limit given our S/N cut is lower in the case of emission line dispersion measurements ($\sim30\ \rm{km\ s^{-1}}$). 

In summary, our stellar kinematic samples (left of Fig. \ref{fig:sample_plot}) provide a good representation of the MaNGA DR17 parent sample outside the low stellar mass regime, while the gas kinematic samples (right of Fig. \ref{fig:sample_plot}) lack a comprehensive coverage of the low-SFR population. These two biases are both reflected in the rotator sample (Fig. \ref{fig:sample_plot_rotators}), which is representative of massive star forming galaxies.

\section{Variations in the STFR: stars $\&$ gas at different radii}
\label{sec:TF_rel}

Throughout this paper, we will analyse the STFR for both stellar and gas rotational velocity, at different radii. For the remainder of this work, we will refer to these relations as the \textbf{stellar} and \textbf{gas} STFRs, noting that in both cases, the independent variable is considered to be stellar mass. The full catalogue of kinematic measurements (rotational velocities and $v/\sigma$ ratios) used in this work is presented in the Supporting Information section (available online only).

In this section, we analyse the effect of using different kinematic tracers (stars and gas) on the STFR. We further test how the computed STFR changes depending on the radius at which the rotational velocity is measured. This comparative analysis is complicated by the different selection effects for the stellar and gas kinematic sub-samples presented in Fig. \ref{fig:sample_plot}. To alleviate this shortcoming, we proceed as follows: In subsection \ref{sec:TF_kinematic_full}, we analyse the differences between stellar and gas STFRs at 1$R_{\rm{e}}$, 1.3$R_{\rm{e}}$ and 2$R_{\rm{e}}$ for the different sub-samples in Fig. \ref{fig:sample_plot}, i.e. without sample matching. This approach is similar to those used in previous STFR studies, whereby results are compared to literature computations for different galaxy samples (e.g. \citealt{bloom_sami_2017}, \citealt{arora_manga_2023}).

Finally, in Section \ref{sec:TF_kinematics_matched} we select the sample of galaxies with stellar and gas kinematics that reach 2$R_{\rm{e}}$, and re-compute the stellar and gas STFRs at 1$R_{\rm{e}}$, 1.3$R_{\rm{e}}$ and 2$R_{\rm{e}}$. This sample selection allows us to interpret the dissimilarities between the stellar and gas STFRs at different radii in a physical manner. We further study the differences in our results with and without sample matching to obtain an informed picture of the biases introduced by comparing STFRs for different galaxy samples.

In this work, we compute the best-fitting STFR using a least-squares (LSQ) fit with intrinsic scatter, implemented using the \texttt{emcee} sampler Python package \citep{foreman-mackey_emcee_2013}. This fit minimises the vertical scatter in velocity, which we aim to study in terms of its physical drivers, while also accounting for the intrinsic scatter in the dependent variable. We take into account the error in velocity when performing the fit, to which we add the uncertainty in the velocity offset $v_{\rm{off}}$ obtained when centering the stellar and gas rotation curves (see Section \ref{sec:rc_extract}). We fit a linear model of the form: 

\begin{equation}
    {\log}(V\ {[ \rm{km\ s^{-1}}  ]}) = a\ {\log}(M_{\star}/\rm{M_{\odot}}) + \textit{b},
	\label{eq:fitting}
\end{equation}

where $a$ and $b$ are the slope and intercept of the STFR, respectively.

For completeness, we re-fit each data set using orthogonal linear regression implemented by the \texttt{HyperFit} package \citep{robotham_hyper-fit_2015}. We present the results of these fits in Appendix \ref{sec:Appendix_Ortho} (Table \ref{tab:appendix_table}), and compare them to the best-fit values from the LSQ fit.

\label{sec:rel_rotators}

\begin{figure*}
	\centering
	\includegraphics[width=\linewidth]{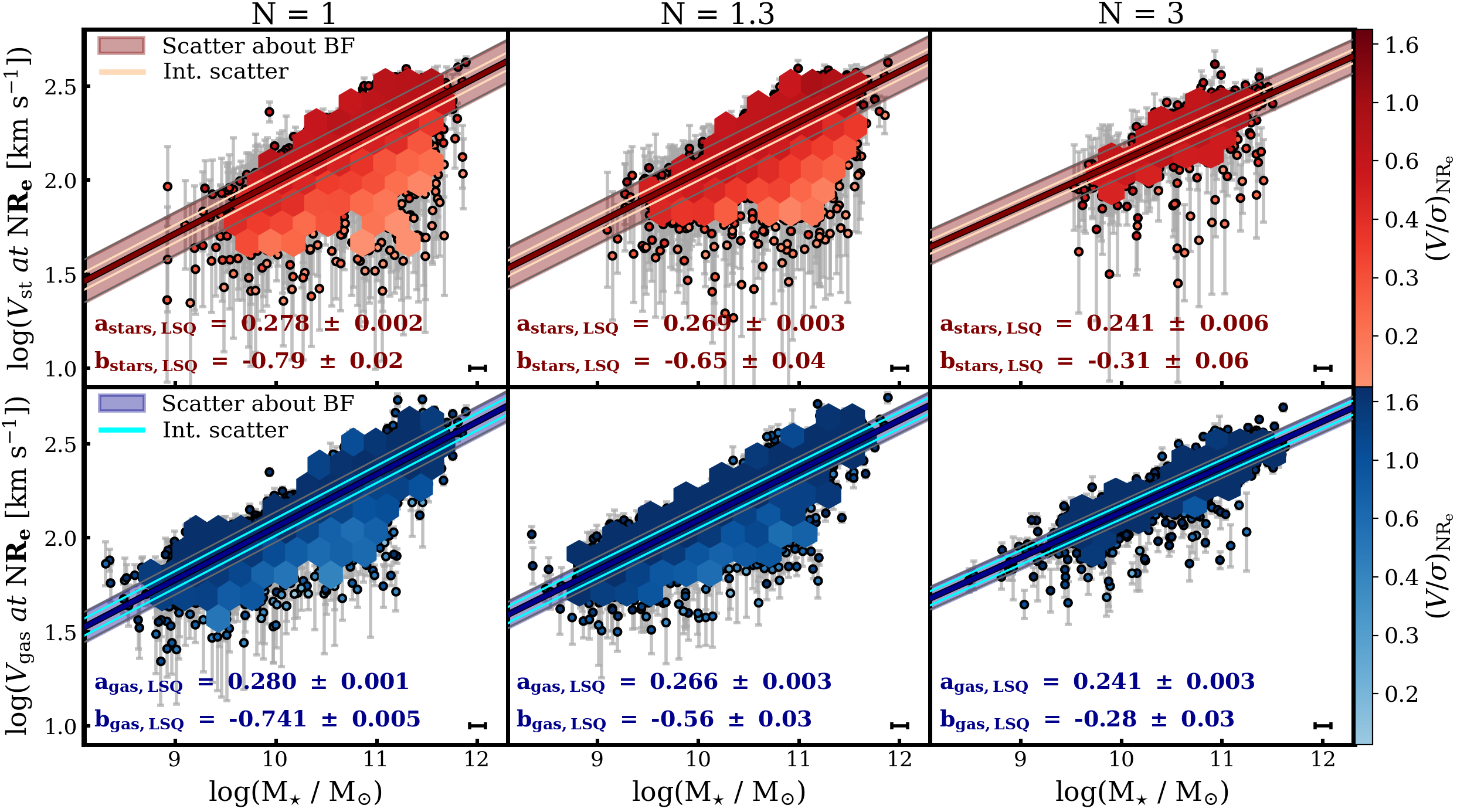}
    \caption{The stellar and gas STFR for our \textbf{stellar} (\textbf{top row}) and \textbf{gas} (\textbf{bottom row}) kinematic samples. The \textbf{left/middle/right} columns show the relation for the velocity at N = (1, 1.3, 2)$\times R_{\rm{e}}$, respectively. Hexagonal bins are plotted for regions with at least 5 data points. The color coding is showing the $(v/\sigma)$ ratio, integrated within the radius at which the velocity is measured. The errors in the velocity measurements are shown in grey, which also account for the uncertainty in the centering of rotation curves in velocity space (see Section \ref{sec:rc_extract}). The dark red/blue lines and shaded regions show the result of a LSQ fit to the data, and the vertical scatter about the best-fit line $s_{\rm{||}}$, respectively. The light red/blue dotted lines show the best-fit intrinsic scatter $s_{\rm{||,INT}}$. The best-fit slope (a) and intercept (b) values from the LSQ fit are displayed on each panel, with their associated uncertainties. The black error bar in the bottom right corner shows the maximum uncertainty in stellar mass for our galaxy sample (see Appendix \ref{sec:Appendix_Ortho}).} 
    \label{fig:TF_kin}
    
\end{figure*}

\definecolor{chromeyellow}{rgb}{1.0, 0.65, 0.0}
\definecolor{caribbeangreen}{rgb}{0.0, 0.8, 0.6}

\renewcommand{\arraystretch}{1.0}
\begin{table*}
\centering
\caption{Compilation of the stellar and gas STFR best-fit parameters from a LSQ fit with intrinsic scatter, for all the galaxy samples used in this work. We show results for our full stellar, gas and common kinematic samples as well as for the selected rotators, at 1$R_{\rm{e}}$, 1.3$R_{\rm{e}}$ and 2$R_{\rm{e}}$. We also present the best-fit relations at 1$R_{\rm{e}}$ and 1.3$R_{\rm{e}}$ only for the common kinematic and rotator samples with stellar and gas kinematics reaching 2$R_{\rm{e}}$. The number of galaxies in each stellar and gas sub-sample at the respective radius is highlighted under the \textcolor{red}{$\rm{N_{gal,st}}$} and \textcolor{blue}{$\rm{N_{gal,g}}$} columns, respectively. The values in square brackets show the uncertainty in the last decimal place of each best-fit value. $s_{\rm{||,INT}}$ is the best-fit (vertical) intrinsic scatter, while $s_{\rm{||}}$ is the vertical scatter about the best-fit line. Uncertainties in $s_{\rm{||}}$ are computed by scattering the velocity values for the respective sample randomly about their uncertainties and re-fitting the STFR; this process is repeated 100 times and the uncertainty in $s_{\rm{||}}$ is the standard error on the mean of the distribution of $s_{\rm{||}}$ values. The mean of this distribution agrees with the $s_{\rm{||}}$ value computed with respect to the fit using the best estimate velocity values within 0.004 dex in all cases. Both $s_{\rm{||,INT}}$ and $s_{\rm{||}}$ values are expressed in dex. The intercept (b) values are expressed in dex, while the slopes (a) are dimensionless. The colored cells correspond to the relations shown in Fig. \ref{fig:TF_kin} (\textcolor{red}{red} and \textcolor{blue}{blue}) and Fig. \ref{fig:TF_rotators} (\textcolor{chromeyellow}{yellow} and \textcolor{caribbeangreen}{green}). }
\setlength\tabcolsep{1.9pt}
\label{tab:breakdown_TF}
\begin{tabular*}{\linewidth}{@{\extracolsep{\fill}} c c|cccc|cccc} 
\multicolumn{1}{c}{\multirow{2}{*}{   Sample   }}                                     & \multirow{2}{*}{   Radius $(\times R_{\rm{e}})$
   | \textcolor{red}{$\rm{N_{gal,st}}$} | \textcolor{blue}{$\rm{N_{gal,g}}$}   }  & \multicolumn{4}{c|}{Stars}                                                                                                                            & \multicolumn{4}{c}{Gas}                                                                                                               \\          
\multicolumn{1}{c}{}                                                            &                               & slope (a)                                               & intercept (b)                                              & $s_{\rm{||,INT}}$       &  $s_{\rm{||}}$                &      slope (a)                                           & intercept (b)                                          & $s_{\rm{||,INT}}$   &   $s_{\rm{||}}$                        \\ 

\hline
\hline
\multirow{6.6}{*}[3em]{\begin{tabular}[c]{@{}c@{}}\\  Kinematic\\(ALL) 
 \end{tabular}}       & $\ \ \ \ \ \ \ \ $ 1   $ \ \ \ \ \ \ \ \ \ $   | $\ $ \textcolor{red}{2683}  | \textcolor{blue}{3430}               & {\cellcolor[rgb]{1,0.373,0.373}}\begin{tabular}{c}\ \ \      0.278[2] \ \ \      \end{tabular} & {\cellcolor[rgb]{1,0.373,0.373}}\begin{tabular}{c} \ \ \    -0.79[2]    \ \ \ \end{tabular} & {\cellcolor[rgb]{1,0.373,0.373}}    \ \ \   0.06  [1]  \  \  & {\cellcolor[rgb]{1,0.373,0.373}} \ \    0.11[3]   \ \ & {\cellcolor[rgb]{0,0.631,1}} \begin{tabular}{c}\ \ \    0.280[1]  \ \ \   \end{tabular} & {\cellcolor[rgb]{0,0.631,1}} \begin{tabular}{c} \ \  \   -0.741[5]  \ \ \  \end{tabular} & {\cellcolor[rgb]{0,0.631,1}}\  \ \   0.04 [1] \ \    & {\cellcolor[rgb]{0,0.631,1}}  \ \   0.08[2] \ \    \\
                                                                                & $\ \ \ \ \ \ \ \ $ 1.3   $ \ \ \ \ \ \  $   | $\ $ \textcolor{red}{2211}  | \textcolor{blue}{3047}                              & {\cellcolor[rgb]{1,0.373,0.373}}\begin{tabular}{c} 0.269[3]\end{tabular}               & {\cellcolor[rgb]{1,0.373,0.373}}\begin{tabular}{c} -0.65[4] \end{tabular}                & {\cellcolor[rgb]{1,0.373,0.373}}0.05[1]   
                                                                                &
                                                                {\cellcolor[rgb]{1,0.373,0.373}}0.11[3]           
                                                                                &  {\cellcolor[rgb]{0,0.631,1}} \begin{tabular}{c} 0.266[3]\end{tabular}                 & {\cellcolor[rgb]{0,0.631,1}}\begin{tabular}{c} -0.56[3]
                                                \end{tabular}                & {\cellcolor[rgb]{0,0.631,1}}0.03[1]    & {\cellcolor[rgb]{0,0.631,1}}0.07[2]             \\
                                                                               & $\ \ \ \ \ \ \ \ $ 2   $ \ \ \ \ \ \ \ \ \ $   | $\ $ \textcolor{red}{530} $\ $  | \textcolor{blue}{1019}                              & {\cellcolor[rgb]{1,0.373,0.373}}\begin{tabular}{c} 0.241[6] \end{tabular}                & {\cellcolor[rgb]{1,0.373,0.373}}\begin{tabular}{c} -0.31[6]
                                                                                \end{tabular}               & {\cellcolor[rgb]{1,0.373,0.373}}0.04[2]   & {\cellcolor[rgb]{1,0.373,0.373}}0.09[4]  
                                                                                
                                                                                & {\cellcolor[rgb]{0,0.631,1}}\begin{tabular}{c} \ 0.241[3]
                                                \end{tabular}                & {\cellcolor[rgb]{0,0.631,1}}\begin{tabular}{c} -0.28[3] \end{tabular}               & {\cellcolor[rgb]{0,0.631,1}}0.03[2] & {\cellcolor[rgb]{0,0.631,1}}0.06[1]      \\

\midrule
                                                            
\multirow{-1.66}{*}{\begin{tabular}[c]{@{}c@{}}\\  Common \\ (stars $\&$ gas) \end{tabular}} & $\ \ \ \ \ \ \ \   $ 1   $ \ \ \ \ \ \ \ \ \ $   | $\ $ \textcolor{red}{1899}  | \textcolor{blue}{1899}                               & \begin{tabular}{c} 0.284[3] \end{tabular}                                                & \begin{tabular}{c} -0.84[4] \end{tabular}                                              & 0.03[2]  & 0.09[3]                                           & \begin{tabular}{c} 0.299[3] \end{tabular}                                            & \begin{tabular}{c} -0.945[7] \end{tabular}                                           & 0.03[2]  & 0.08[2]                                       \\
                                                                                & $\ \ \ \ \ \ \ \ $ 1.3   $\ \ \  \ \ \  $   | $\ $ \textcolor{red}{1458}  | \textcolor{blue}{1458}                            & \begin{tabular}{c} 0.280[4]\end{tabular}                                                & \begin{tabular}{c} -0.76[5] \end{tabular}                                               & 0.03[3]                & 0.08[3]                              & \begin{tabular}{c} 0.290[4]\end{tabular}                                            & \begin{tabular}{c} -0.82[6] \end{tabular}                                           & 0.03[2]           & 0.07[2]                         \\

                                                                                & $\ \ \ \ \ \  $ 2   $ \ \ \  \ \ \ \ \ \  $   | $\ $ \textcolor{red}{235} $\ $  | \textcolor{blue}{235}                              & \begin{tabular}{c} 0.26[1] \end{tabular}                                                & \begin{tabular}{c} -0.48[1] \end{tabular}                                               & 0.02[4]              & 0.07[4]                              &  \begin{tabular}{c} 0.264[1] \end{tabular}                                            & \begin{tabular}{c} -0.51[4]\end{tabular}                                           & 0.02[3]                     & 0.06[3]                             \\
                                                            \midrule

\multirow{6.22}{*}[3em]{\begin{tabular}[c]{@{}c@{}}Common  \\ w/ $R_{\rm{max}} \geqslant 2R_{\rm{e}}$ \\  \end{tabular}}   & $\ \ \ \ \ \  $ 1   $\ \  \ \ \ \ \ \  \  $   | $\ $ \textcolor{red}{235} $\ $  |  \textcolor{blue}{235}                              & \begin{tabular}{c} 0.310[8] \end{tabular}                                                & \begin{tabular}{c} -1.13[6] \\ \end{tabular}                                              & 0.02[1]    & 0.08[3]                                       & \begin{tabular}{c} 0.295[2] \end{tabular}                                            & \begin{tabular}{c} -0.90[2] \end{tabular}                                           & 0.03[1]    & 0.07[2]                                    \\
                                                                                & $\ \ \ \ \ \  $ 1.3   $\ \ \ \ \ \    $   | $\ $ \textcolor{red}{235} $\ $  |  \textcolor{blue}{235}                                                       & \begin{tabular}{c} 0.294[7]
                                                                                \end{tabular}                                                & \begin{tabular}{c} -0.90[9] \end{tabular}                                               & 0.02[2]                   & 0.06[3]                        & \begin{tabular}{c} 0.276[2]\end{tabular}                                            & \begin{tabular}{c} -0.67[2] \end{tabular}                                           & 0.02[1]           & 0.06[3]                           \\
                                                              
\midrule 
                                                           
\multirow{6.24}{*}[3em]{\begin{tabular}[c]{@{}c@{}}\\ Rotators \end{tabular}}                                                     & $\ \ \ \ \ \  $ 1   $\ \ \ \ \ \ \ \  \  $   | $\ $ \textcolor{red}{879} $\ $  |  \textcolor{blue}{879}                                                         & {\cellcolor[rgb]{1,0.859,0.216}}\begin{tabular}{c} 0.282[3] \end{tabular}                & {\cellcolor[rgb]{1,0.859,0.216}}   \begin{tabular}{c} -0.78[3] \end{tabular}               & {\cellcolor[rgb]{1,0.859,0.216}}  0.02[2]  & {\cellcolor[rgb]{1,0.859,0.216}}  0.07[3]  & {\cellcolor[rgb]{0,0.898,0.62}} \begin{tabular}{c} 0.282[2] \end{tabular}             & {\cellcolor[rgb]{0,0.898,0.62}}  \begin{tabular}{c} -0.75[4] \end{tabular}            & {\cellcolor[rgb]{0,0.898,0.62}}  0.02[2] & {\cellcolor[rgb]{0,0.898,0.62}}  0.06[3]          \\
                                                                                & $\ \ \ \ \ \  $ 1.3   $\ \ \ \ \ \   $   | $\ $ \textcolor{red}{644} $\ $  |  \textcolor{blue}{644}                                                       & {\cellcolor[rgb]{1,0.859,0.216}}\begin{tabular}{c}  0.279[2]\end{tabular}                & {\cellcolor[rgb]{1,0.859,0.216}} \begin{tabular}{c} -0.73[3] \end{tabular}               & {\cellcolor[rgb]{1,0.859,0.216}} 0.02[3]    
                                                                      & {\cellcolor[rgb]{1,0.859,0.216}}  0.06[3]          
                                                                                & 
                                                                                
                                                                                {\cellcolor[rgb]{0,0.898,0.62}}\begin{tabular}{c} \ 0.275[2]\end{tabular}             & {\cellcolor[rgb]{0,0.898,0.62}}\begin{tabular}{c} \ -0.65[3]\end{tabular}            & {\cellcolor[rgb]{0,0.898,0.62}}\ 0.02[2]
                                                                                  & {\cellcolor[rgb]{0,0.898,0.62}}\ 0.06[3]
                                                                                \\
                                                                                & $\ \ \ \ \ \  $ 2   $\ \ \ \    \ \  \ \ \  $   | $\ \  $ \textcolor{red}{79} $\  \ $  | $ \ $ \textcolor{blue}{79}                                                         & {\cellcolor[rgb]{1,0.859,0.216}} \begin{tabular}{c} 0.27[2] \end{tabular}                & {\cellcolor[rgb]{1,0.859,0.216}}\begin{tabular}{c} \ -0.60[2] \end{tabular}               & {\cellcolor[rgb]{1,0.859,0.216}}\ 0.01[3]    
                                                                       & {\cellcolor[rgb]{1,0.859,0.216}}\ 0.05[4]           
                                                                                & {\cellcolor[rgb]{0,0.898,0.62}}\begin{tabular}{c} 0.26[2]\end{tabular}             & {\cellcolor[rgb]{0,0.898,0.62}}\begin{tabular}{c} \ -0.48[3] \end{tabular}            &  

                                                                     {\cellcolor[rgb]{0,0.898,0.62}}\begin{tabular}{c} \  0.01[3] \end{tabular}            
                                                                      & {\cellcolor[rgb]{0,0.898,0.62}}\ 0.04[4]           

\\ 

\midrule 

\multirow{6.2}{*}[3em]{\begin{tabular}[c]{@{}c@{}}Rotators   \\ w/ $R_{\rm{max}} \geqslant 2R_{\rm{e}}$ \end{tabular}}                                                      & $\ \ \ \ \ \  $ 1   $\ \ \ \    \ \  \ \ \  $   | $\ \  $ \textcolor{red}{79} $\  \ $  | $\ $ \textcolor{blue}{79}                             & \begin{tabular}{c}\ 0.28[2]\end{tabular}                &    \begin{tabular}{c} \ -0.82[2] \end{tabular}               &  \ 0.02[2] &  \ 0.06[2] &  \begin{tabular}{c} 0.26[2]\end{tabular}             &   \begin{tabular}{c} \ -0.55[9] \end{tabular}            & \ \ 0.02[1]   &   \ 0.06[2]    \\
                                                                                & $\ \ \ \ \ \  $ 1.3   $\ \ \ \    \ \      $   | $\ \  $ \textcolor{red}{79} $\  \ $  | $\ $ \textcolor{blue}{79}                           & \begin{tabular}{c} \ 0.28[2] \end{tabular}                &  \begin{tabular}{c} \ -0.8[1] \end{tabular}               & \ 0.01[3]   
                                                                             & \ 0.05[3]      
                                                                                & \begin{tabular}{c} 0.26[3]\end{tabular}             & \begin{tabular}{c} \ -0.48[5]\end{tabular}            & \ 0.01[2]   
                                                                           & \ 0.06[2]       
                                                                                \\

\midrule 
\midrule

\end{tabular*}
\end{table*}

\begin{figure*}
	\centering
	\includegraphics[width=\linewidth]{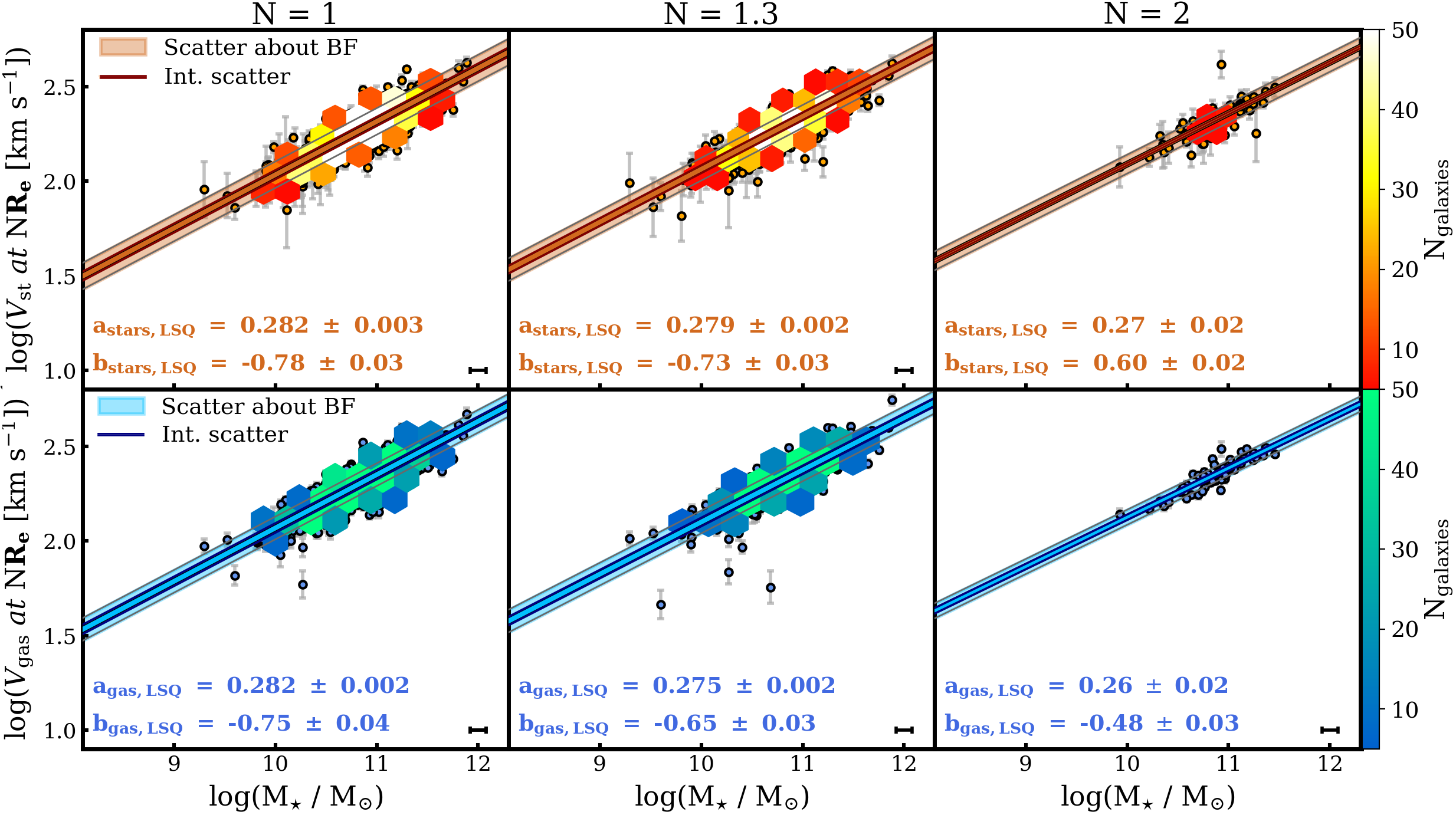}
    \caption{The \textbf{stellar (top row)} and \textbf{gas (bottom row)} STFRs for our \textbf{rotator} sample, in the same format as Fig. \ref{fig:TF_kin}. }
    \label{fig:TF_rotators}
\end{figure*}

\subsection{Stellar $\&$ gas STFRs at different radii without sample matching}
\label{sec:TF_kinematic_full}

We present in Fig. 3 the stellar and gas STFRs for our MaNGA kinematic samples. The dark red and blue solid lines show the result of the LSQ fit. The shaded region and light red $\&$ blue lines show the vertical scatter about the best-fitting relation $s_{||}$ (computed as half the difference between the $84^{\rm{th}}$ and $16^{\rm{th}}$ percentiles of the offset distribution from the best-fit STFR) and the best-fit intrinsic scatter in the relations $s_{\rm{||,INT}}$, respectively. A full compilation of the best fitting  parameters for the various kinematic samples used in this work is given in Table \ref{tab:breakdown_TF}.

We make the following observations when analysing the stellar and gas STFRs \textit{individually}: The STFRs for both baryonic tracers become shallower between 1$R_{\rm{e}}$ and 2$R_{\rm{e}}$ (the slope $a$ is decreasing). The intercept $b$ of the STFR for both tracers is also becoming higher as the radius used to measure the velocity is increasing. The largest differences are reported between the relations at 1.3$R_{\rm{e}}$ and 2$R_{\rm{e}}$, as displayed on Fig. \ref{fig:TF_kin}. The vertical scatter in the two relations $s_{\rm{||}}$ is decreasing with radius used to probe the velocity (0.11, 0.11, 0.09 dex for stars, and 0.08, 0.07, 0.06 dex for gas at 1$R_{\rm{e}}$, 1.3$R_{\rm{e}}$, 2$R_{\rm{e}}$). The same trend is valid when comparing intrinsic scatters ($s_{\rm{||,INT}}$), as outlined in Table \ref{tab:breakdown_TF} (0.06 - 0.04 dex for stars and 0.04 - 0.03 dex for gas).

When comparing the STFRs of stars and gas at each radius, we note that
the slopes of the two relations are consistent within errors at 1$R_{\rm{e}}$ ($a_{\rm{stars}} = 0.278 \pm 0.002$ vs $a_{\rm{gas}} = 0.280 \pm 0.001$), 1.3$R_{\rm{e}}$ ($a_{\rm{stars}} = 0.269 \pm 0.003$ vs $a_{\rm{gas}} = 0.266 \pm 0.003$), and 2$R_{\rm{e}}$ ($a_{\rm{stars}} = 0.241 \pm 0.006$ vs $a_{\rm{gas}} = 0.241 \pm 0.003$). The gas relation is exhibiting slightly less vertical scatter about the best-fit line at all probed radii (0.06 - 0.08 dex) compared to the stellar relation (0.09 - 0.11 dex). The same result is recovered when comparing the intrinsic scatters in the relations, with $s_{\rm{||,INT}}$ in the stellar STFR being larger by 0.02 dex than in the gas at 1$R_{\rm{e}}$ and 1.3$R_{\rm{e}}$, and by 0.01 dex at 2$R_{\rm{e}}$.
The slope decrease and intercept increase for both stellar and gas relations between 1$R_{\rm{e}}$ and 2$R_{\rm{e}}$ are qualitatively consistent with the findings of \cite{yegorova_radial_2007}. Our study confirms the existence of a family of independent STFRs up to 2$R_{\rm{e}}$, for both stellar and gas rotation, in accord with the presence of a dark matter component exerting an influence on the kinematics of both stars and gas that increases with radius (\citealt{yegorova_radial_2007}).

The slightly larger intrinsic scatter in the stellar STFR at 1$R_{\rm{e}}$ and 1.3$R_{\rm{e}}$ compared to the gas (when the full stellar and gas kinematic samples are considered) is easily explainable by the variation in kinematic morphology for our sample. This variation is shown by the color coding in Fig. \ref{fig:TF_kin} that highlights a clear trend between vertical scatter and $v/\sigma$ ratio within the respective radius for stellar kinematics. While the same trend is present in the gas relation (i.e. galaxies with more turbulent gas discs being preferentially found below the STFR) the range of $v/\sigma$ for the gas is lower than for the stars. This difference in $v/\sigma$ range is due to the collisionless nature of stars that allows them to form into stable dispersion-supported structures (classical bulges). Galaxies hosting such structures are expected to lie below the SFMS (\citealt{fraser-mckelvie_sami_2021}) at a given stellar mass, and are thus expected to be found in the stellar kinematic sample, but not the gas one (Fig. \ref{fig:sample_plot}). On the other hand,  dispersive motions in the gas are reduced due to viscous friction forces. At 2$R_{\rm{e}}$ where the intrinsic scatters are consistent within uncertainties, both our stellar and gas kinematic samples show little variation in their kinematic morphology, with $(v/\sigma)_{2R_{\rm{e}}}\ \gtrsim  1$ for most galaxies.

We present the stellar and gas STFRs for our sample of rotators (Section \ref{sec:final_sample}) in Fig. \ref{fig:TF_rotators}. In the case of the stellar relation (top row), we find no significant variation between 1$R_{\rm{e}}$ and 1.3$R_{\rm{e}}$ in either slope ($\Delta a \leqslant 0.03$), intercept ($\Delta b \leqslant 0.05 $ dex) or intrinsic scatter ($\Delta s_{\rm{||,INT}} \approx 0$), and a relation consistent with that present for the full kinematic sample at 1$R_{\rm{e}}$. The stellar rotator STFR becomes slightly shallower at $2R_{\rm{e}}$, albeit with a slope still consistent with that of the relations in the inner parts (1$R_{\rm{e}}$ and 1.3$R_{\rm{e}}$). The gas STFR for rotators is consistent with that of the stars at 1$R_{\rm{e}}$ within uncertainties, but becomes slightly shallower at 1.3$R_{\rm{e}}$ and 2$R_{\rm{e}}$, as shown by a decrease in slope (bottom row of Fig. \ref{fig:TF_rotators}). Differences between stellar and gas relations for rotators at different radii are within 0.1 dex in log(velocity) for the entire range of stellar mass probed.

We look for variations in the STFR vertical scatter with stellar mass by splitting our individual sub-samples in Fig. \ref{fig:TF_kin} and Fig. \ref{fig:TF_rotators} into 4 bins of equal width in log$(M_{\star}/\rm{M_{\odot}})$. The largest variations of $s_{\rm{||}}$ are found for the stellar relations at 1$R_{\rm{e}}$ and 1.3$R_{\rm{e}}$, and are 0.06 dex from the median at both radii. These variations are both found for the highest stellar mass bins ($10^{11.1} - 10^{11.9} \rm{M_{\odot}}$ and $10^{11.2} - 10^{11.9} \rm{M_{\odot}}$ at 1$R_{\rm{e}}$ and 1.3$R_{\rm{e}}$, respectively), and are significant (the scatters scaled by the median velocity uncertainties in each of the two respective bins are a factor of 4.0 and 3.6 at 1$R_{\rm{e}}$ and 1.3$R_{\rm{e}}$, respectively). This result is in agreement with the findings of \cite{fraser-mckelvie_beyond_2022}, who reported that slow rotators only contribute significantly to the mass budget of MaGNA DR17 galaxies above log($M_{\star} / \rm{M_{\odot}}$) = 11.25. As such, we obtain a larger spread in rotational-to-dispersional support (and scatter) in the higher stellar mass bins for the stellar STFR, as the slow-rotator population is well sampled in these mass bins. We note that, outside of these cases, the deviations from the median $s_{\rm{||}}$ for the sub-samples in Fig. \ref{fig:TF_kin} and Fig. \ref{fig:TF_rotators} are <0.02 dex across each respective stellar mass range.

\begin{figure}
	\centering
	\includegraphics[width=\columnwidth]{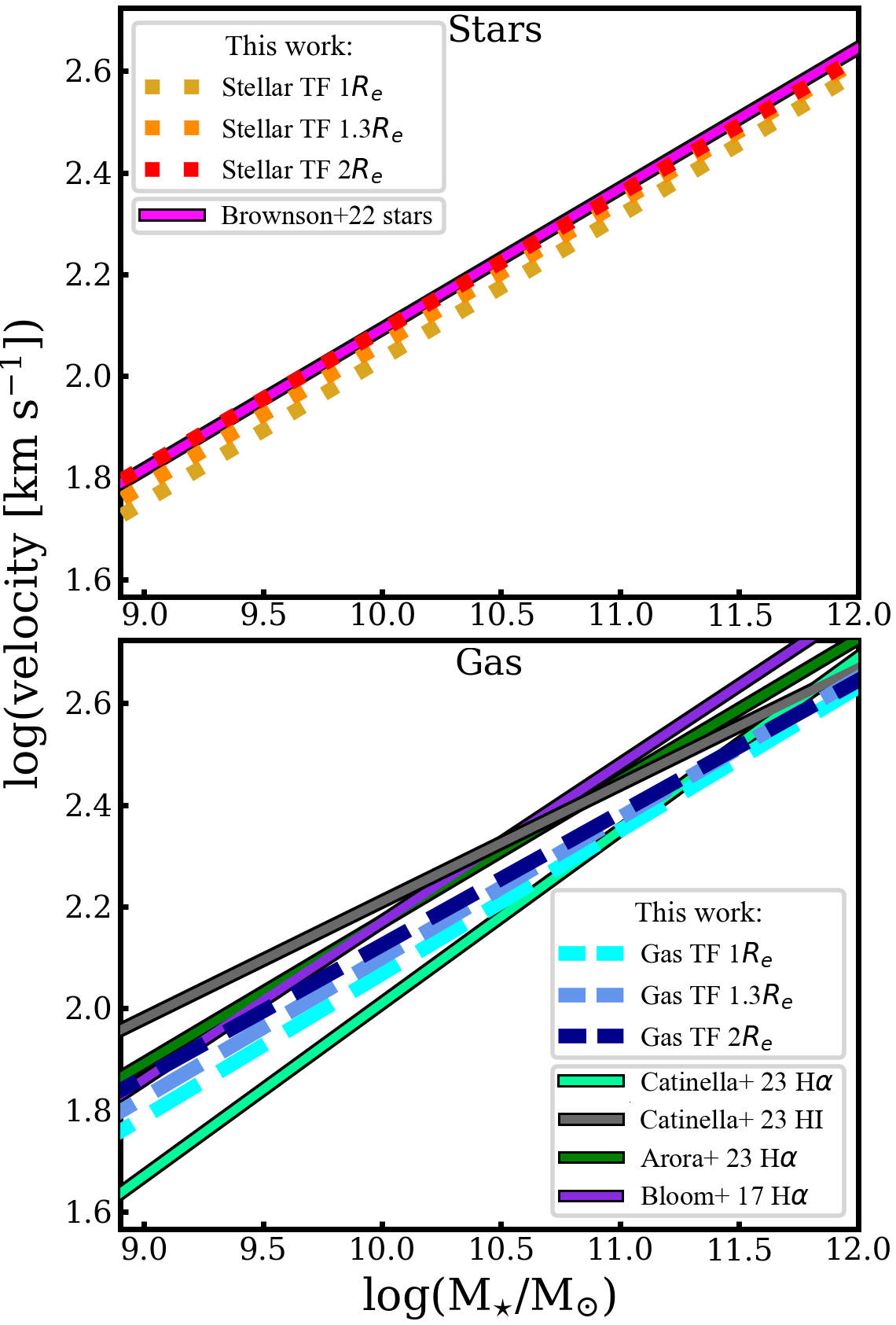}
    \caption{A comparison of STFRs for the rotator samples in this study with literature computations. \textbf{Top:} The stellar STFRs at 1$R_{\rm{e}}$, 1.3$R_{\rm{e}}$ and 2$R_{\rm{e}}$ for our rotator samples (dotted lines) in comparison with the relation computed by \protect\cite{brownson_what_2022}. \textbf{Bottom:} A comparison between the gas STFRs at 1$R_{\rm{e}}$, 1.3$R_{\rm{e}}$ and 2$R_{\rm{e}}$ for rotators presented in this work (dashed lines), and various literature sources, as indicated in the legend.}
    \label{fig:TF_comparison_samples}
\end{figure}

In Fig. \ref{fig:TF_comparison_samples}, we present a comparison of our stellar and gas rotator STFRs with various literature STFR computations, with stars (top), H$\rm{\alpha}$ and \ion{H}{I} (bottom) being used as kinematic tracers. We first note a general agreement of our results with all of the comparison relations, with differences in log(velocity) within 0.19 dex over the stellar mass range probed by our samples. Our stellar STFR for rotators agrees remarkably well with that of \cite{simcha_brownson_what_2022}. Differences in log(velocity) are <0.02 dex compared to our stellar STFR at 2$R_{\rm{e}}$, corresponding to a maximum ratio in velocity of 1.04. The spread in relations based on gas rotation is larger, with stronger divergences at lower stellar masses (< $10^{10} \rm{M_{\odot}}$). This discrepancy may be attributed to different methods of computing stellar masses, e.g. \textit{i}-band magnitudes and \textit{g-i} colour in \cite{bloom_sami_2017} and \cite{catinella_sami-h_2023}, or \textit{g-r} and \textit{g-z} optical colors and luminosities in \cite{arora_manga_2023}. Such discrepancies are expected to be more pronounced in the low-$M_{\star}$ regime. We also note that the selection criteria for our rotator sample (Section \ref{sec:final_sample}) are expected to exclude kinematically disturbed systems. Such objects have been shown to predominantly scatter below the gas STFR in \cite{bloom_sami_2017} and \cite{catinella_sami-h_2023}, thus increasing the discrepancies between our work and these studies. Differences from our gas STFR for rotators at 2$R_{\rm{e}}$ are within 0.13, 0.20, 0.12, 0.08 dex in log(velocity) for \cite{bloom_sami_2017}, \cite{catinella_sami-h_2023}(H$\rm{\alpha}$), \cite{catinella_sami-h_2023}(\ion{H}{I}) and \cite{arora_manga_2023} respectively, corresponding to a maximum ratio in velocity of 1.35, 1.58, 1.31, 1.19.
 
Differences in the gas relations may also be contributed to by different radii used to estimate rotational velocities: 2.2$R_{\rm{e}}$ and 1.3$R_{\rm{e}}$ for \cite{bloom_sami_2017} and H$\rm{\alpha}$ measures in \cite{catinella_sami-h_2023} respectively (extrapolated if required), and the radius at which the surface brightness is equal to 23.5 mag $\rm{arcsec^{-2}}$ for \cite{arora_manga_2023}. The differences between our gas STFR and that of \cite{catinella_sami-h_2023} computed for \ion{H}{I} rotation (more pronounced at low stellar masses) can largely be explained by aperture effects, whereby our H$\rm{\alpha}$ kinematic measurements do not reach the flat part of rotation curves in dwarf galaxies.
Furthermore, differences in the gas relations of Fig. \ref{fig:TF_comparison_samples} can also be attributed to variations in the physical properties of the galaxy samples used in each study: a morphologically limited sample of massive late-type galaxies for \cite{arora_manga_2023}, a sample of predominantly star forming systems with specific SFRs above $10^{-11}\ \rm{yr^{-1}}$ for \cite{catinella_sami-h_2023}, and a sample of varied optical morphology, but biased towards late-types for \cite{bloom_sami_2017}.

The spread in log(velocity) for rotator STFRs computed in this work for the same kinematic tracer at different radii (<0.06 dex for stars and <0.07 dex for gas) is significantly lower than the spread in relations from various literature sources (within 0.19 dex for our probed range of stellar masses, considering the relations for both stars and gas), which are representative of different galaxy samples as discussed. This finding suggests that differences in sample selection dominate over systematic uncertainties introduced by differences in radius at which the velocity is measured. 

In summary, we report a variation of the STFR with radius for both the stellar and gas kinematic samples, such that relations at larger radii are shallower for both baryonic components. The stellar and gas STFRs have slopes that are consistent at all probed radii (Fig. \ref{fig:TF_kin}). 
When selecting only rotationally dominated galaxies, the stellar STFR remains constant for velocities at 1$R_{\rm{e}}$ and 1.3$R_{\rm{e}}$, and becomes slightly shallower at 2$R_{\rm{e}}$. The gas STFR 
becomes shallower as the radius used to estimate the velocity increases (Fig. \ref{fig:TF_rotators}). 
We find a general agreement of our relations for rotators with previous studies of the stellar and gas STFRs, although with slightly larger differences between the gas relations below $10^{10} \rm{M_{\odot}}$ (Fig. \ref{fig:TF_comparison_samples}).
In the following subsection, we present the stellar and gas STFRs after performing sample matching, i.e. for the same sample of galaxies with kinematics for both baryonic components that reach 2$R_{\rm{e}}$.

\subsection{Stellar $\&$ gas STFRs at different radii for the same galaxy sample}
\label{sec:TF_kinematics_matched}

\begin{figure*}
	\centering
	\includegraphics[width=\linewidth]{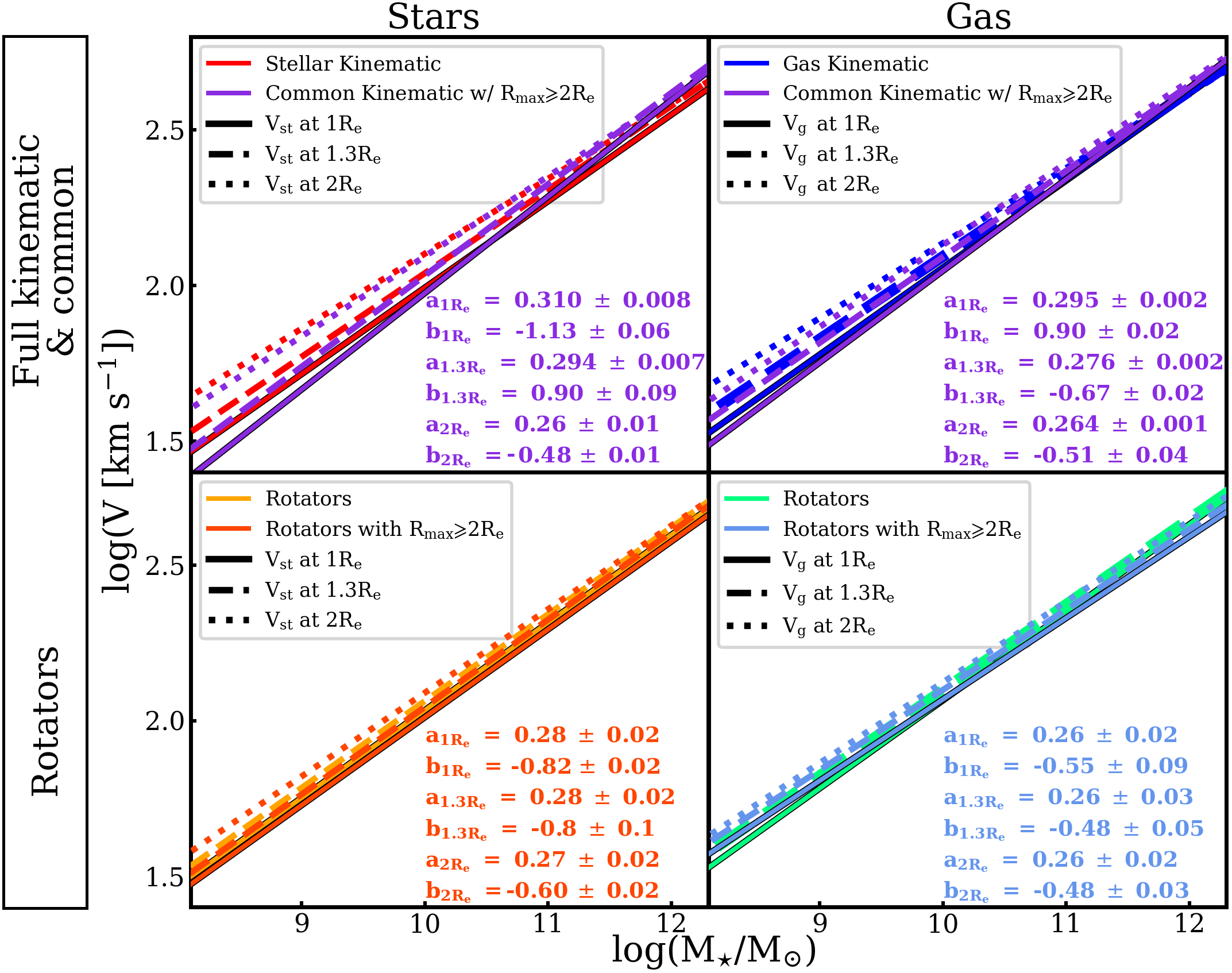}
    \caption{\textbf{Top: }A compilation of the differences in stellar (\textbf{left}) and gas (\textbf{right}) STFRs at 1$R_{\rm{e}}$, 1.3$R_{\rm{e}}$  and 2$R_{\rm{e}}$  between using the full kinematic samples (red/blue) and using only the sub-samples with both stellar and gas kinematics reaching 2$R_{\rm{e}}$ (purple). \textbf{Bottom:} The differences in stellar and gas STFR between using the full rotator samples at each radius  (yellow and green lines, respectively) and only the rotator sub-samples that reach 2$R_{\rm{e}}$ (orange and light blue lines, respectively). As such, the purple lines (top row) and the orange/blue lines (bottom row) reflect the STFRs for stars and gas at different radii, for the same sample of galaxies (only selecting rotators in the case of the bottom row), and are directly comparable (differences are not driven by sample biases). We show the best-fit slope (a) and intercept (b) corresponding to these relations on each panel, matched in color to the respective lines, and with the subscript indicating the radius at which the velocity is measured. }
    \label{fig:TF_comparison_sample_match}
\end{figure*}

The results presented in the above section are obtained using different samples for stars and gas. As such, we need to assess whether different selection criteria may impact any of the conclusions presented.
We select the common kinematic sample with stellar and gas kinematics reaching 2$R_{\rm{e}}$ (Section \ref{sec:final_sample}) and refit the stellar and gas STFRs at 1$R_{\rm{e}}$, 1.3$R_{\rm{e}}$ and 2$R_{\rm{e}}$ using the functional form of equation \ref{eq:fitting}. We also repeat the analysis in the section above for the sub-sample of rotators with kinematics reaching 2$R_{\rm{e}}$. Fig. \ref{fig:TF_comparison_sample_match} shows a compilation of these relations, in comparison with the STFRs for the whole kinematic and rotator samples (presented in Fig. \ref{fig:TF_kin} and Fig. \ref{fig:TF_rotators})

In the case of the stellar STFRs (top left panel of Fig. \ref{fig:TF_comparison_sample_match}), we note that sample matching for the kinematic sample (purple versus red lines) produces a more pronounced difference in the lower end of the probed stellar mass range (below $\sim 10^{10}\ \rm{M_{\odot}}$) such that the relation becomes steeper when the same sample of galaxies is considered. This difference, while within the scatter about the best-fit relation (below 0.07 dex in log(velocity), corresponding to a maximum ratio in velocity of 1.17) at all radii, highlights the contribution of the more gas-poor or dispersion-supported galaxies not considered in the matched sample, which are predominantly found below the stellar STFR (Fig. \ref{fig:TF_kin}). We find a similar result when considering the rotator galaxies that reach 2$R_{\rm{e}}$ (bottom left panel of Fig. \ref{fig:TF_comparison_sample_match}), although with only very small differences  (<0.01 dex in log (velocity), corresponding to a maximum ratio in velocity of 1.02) from the relations for the full rotator samples at the same radius.

When comparing the gas STFRs with and without sample matching (top right of Fig. \ref{fig:TF_comparison_sample_match}), we find similar trends as for the stars (relations in the sample-matched case are steeper than when the full kinematic sample is considered), although the differences are smaller (below 0.03 dex in log(velocity), corresponding to a maximum ratio in velocity of 1.08) at all radii. When only rotators are considered (bottom right of Fig. \ref{fig:TF_comparison_sample_match}), no radial variations are noted and the relations are consistent within uncertainties (between the sample-matched and unmatched cases). This finding  highlights that the radial variations in the gas STFRs in Fig. \ref{fig:TF_kin} and \ref{fig:TF_rotators} are dominated by the requirement of having H$\rm{\alpha}$ emission traceable to sufficiently high radii, and the further requirement of having high-enough stellar continuum does not affect the results.

The relations shown in purple in the top row and orange/light blue in the bottom row of Fig. \ref{fig:TF_comparison_sample_match} are representative of the same galaxy sample and thus physically comparable between different radii and kinematic tracers. 
The stellar and gas STFRs become slightly shallower at larger radii (as indicated by the best-fit slope values displayed on Fig. \ref{fig:TF_comparison_sample_match}), indicative of the change in the gravitational potential between 1$R_{\rm{e}}$ and 2$R_{\rm{e}}$ (same trends as in Fig. \ref{fig:TF_kin}). 
When comparing the stellar and gas STFRs, a steeper relation for stars at 1$R_{\rm{e}}$ and 1.3$R_{\rm{e}}$ compared to the gas is evident, while the slopes of the two relations are consistent wihtin uncertainties at 2$R_{\rm{e}}$. 

This finding suggests that, statistically, the two baryonic components are in different states of dynamical equilibrium in the inner parts ($\leqslant$1.3$R_{\rm{e}}$). We note however that the gas STFR, while shallower, is above the stellar one for the entire mass range probed by our galaxy samples (shown in Fig. \ref{fig:sample_plot}) at 1$R_{\rm{e}}$ and 1.3$R_{\rm{e}}$. The gas component is collisional and thus affected by viscous forces that are not acting on the stellar population. As such, while the gas may be subject to turbulent motions, its dynamical support will be dominated by rotation (as opposed to dispersion). The stars, on the other hand, can form dispersion-supported structures (classical bulges) in the inner parts of galaxies, in which random dispersive motions can dominate the dynamical support, thus explaining the stellar STFR being below the gas one at 1$R_{\rm{e}}$ and 1.3$R_{\rm{e}}$ (with the note that such differences are decreasing with increasing stellar mass).
Such discrepancies are however not present at 2$R_{\rm{e}}$, suggesting that in the outer edges both components are in the same state of dynamical equilibrium, when the sample of galaxies with both stellar and gas kinematics is considered. We note, however, the caveat of the kinematic sample used here not being representative of the low (<$10^{9.5}\ \rm{M_{\odot}}$) stellar masses (Fig. \ref{fig:sample_plot}). This result is indicative of an overall inside-out formation scenario for galaxies in the nearby Universe, whereby stellar populations in the galactic centres form first and are dynamically hotter than in the outskirts.

The vertical scatter $s_{||}$ in the stellar STFR computed for the common kinematic sample reaching 2$R_{\rm{e}}$ shows only small (<0.02 dex) variations between the probed radii: 0.08, 0.06 and 0.07 dex at 1$R_{\rm{e}}$, 1.3$R_{\rm{e}}$ and 2$R_{\rm{e}}$, respectively. This scatter is similar to that in the gas relation for the same sample (0.07, 0.06 and 0.06 dex at 1$R_{\rm{e}}$, 1.3$R_{\rm{e}}$ and 2$R_{\rm{e}}$, respectively). The intrinsic scatter in these relations is approximately constant between the probed radii, and consistent within uncertainties between the relations for stars (0.02 dex at all radii) and gas (0.03 dex at 1$R_{\rm{e}}$ and 0.02 dex at 1.3$R_{\rm{e}}$ and 2$R_{\rm{e}}$). This result highlights that the larger intrinsic scatter in the stellar STFR compared to the gas one for the entire kinematic sample (Fig. \ref{fig:TF_kin}) is the result of a different sample selection function.

When analysing the stellar STFR variation for our rotator sample with kinematics reaching 2$R_{\rm{e}}$ (bottom left panel of Fig. \ref{fig:TF_comparison_sample_match}), there is no variation between 1$R_{\rm{e}}$ and 1.3$R_{\rm{e}}$. The relation at 2$R_{\rm{e}}$ is slightly shallower compared to the inner parts, albeit having a slope consistent within uncertainties with those at 1$R_{\rm{e}}$ and 1.3$R_{\rm{e}}$ ($a_{2R_{\rm{e}}}=0.27 \pm 0.02$ compared to $a_{1.3R_{\rm{e}}}=0.28 \pm 0.02$ and $a_{1R_{\rm{e}}}=0.28 \pm 0.02$). The same is valid for the gas STFR for rotators (bottom right panel of Fig. \ref{fig:TF_comparison_sample_match}). We recover a slightly shallower gas STFR for rotators compared to the stellar relation (as shown by the best-fit slopes displayed on the bottom row of Fig. \ref{fig:TF_comparison_sample_match}), albeit still consistent within uncertainties at all radii. This result highlights a statistically similar state of dynamical equilibrium for both kinematic tracers up to 2$R_{\rm{e}}$, when only rotationally dominated galaxies are considered.

In summary, sample matching (considering the common kinematic and rotator samples reaching 2$R_{\rm{e}}$) produces a slightly steeper stellar STFR (compared to the case with no sample matching), whereas the gas STFR remains largely unchanged (Fig. \ref{fig:TF_comparison_sample_match}). The radial trends between 1$R_{\rm{e}}$ and 2$R_{\rm{e}}$ present for the entire kinematic sample (Section \ref{sec:TF_kinematic_full}) are maintained in the sample matched case. A comparison between the stellar and gas relations reveals a steeper stellar STFR at 1$R_{\rm{e}}$ and 1.3$R_{\rm{e}}$ than the gas one, while the two are comparable in the outer edges (2$R_{\rm{e}}$) Finally, the scatter in the stellar and gas STFRs are consistent within 0.01 dex at all radii when sample matching is performed.

\begin{figure*}
	\centering
	\includegraphics[width=\linewidth]{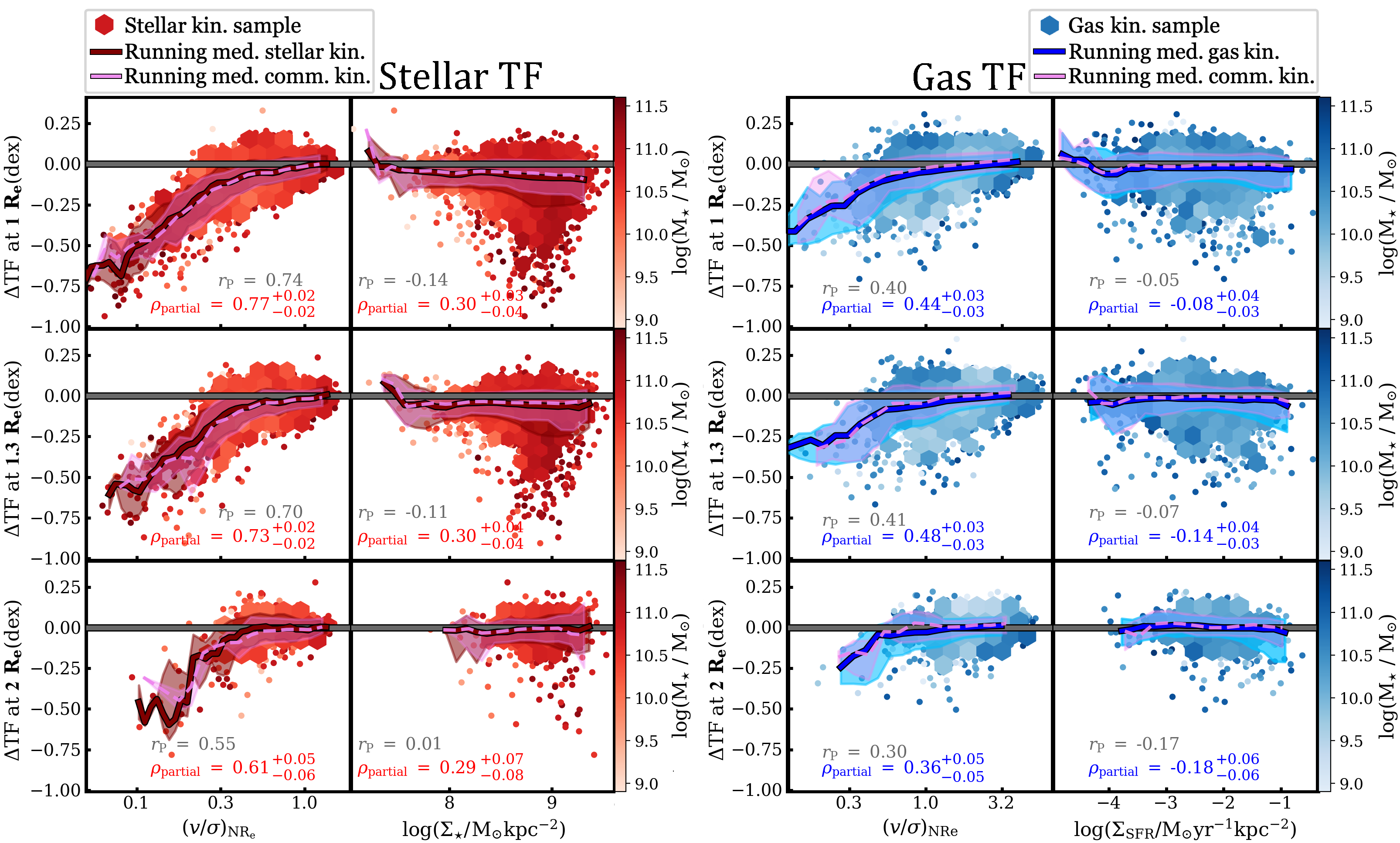}
    \caption{Dependence of \textcolor{red}{stellar} (\textbf{left columns}) and \textcolor{blue}{gas} (\textbf{right columns}) STFRs vertical scatter from a LSQ fit ($\Delta$TF) on:  integrated log$(v/\sigma)$ and log$(\Sigma_{\star})$ for stars;  integrated log$(v/\sigma)$ and log$(\Sigma_{\rm{SFR}})$ for gas. In the case of log$(v/\sigma)$, the parameter axis values show the corresponding non-logged values. The \textbf{top/middle/bottom} rows show the residual dependence for the STFR computed using velocities at \textbf{1, 1.3 and 2}$R_{\rm{e}}$, respectively. The red/blue hexagonal bins and circles show the distribution for the \textit{stellar/gas kinematic samples} (bins are plotted for regions with at least 5 data points). The color coding shows the stellar mass of the respective galaxies. The dark red/blue lines and shaded regions show the running medians for the stellar/gas kinematic samples, and the $\rm{16^{th}}$ - $\rm{84^{th}}$ percentile intervals. The partial linear correlation coefficients (see Section \ref{sec:scatter physical props})  between $\Delta \rm{TF}$ at the respective radius, and the parameter on the x-axis, is shown on each panel in red or blue. We also show the absolute linear correlation coefficient $r_{\rm{P}}$ in grey, for comparison. The running median for the \textit{common kinematic sample} is shown by the pink dashed line for comparison. The horizontal grey line is placed at $\Delta \rm{TF}$ = 0. The uncertainties in $\rho_{\rm{partial }}$ show the 5 per cent and 95 per cent parametric confidence intervals. }
    \label{fig:dBF_both}
\end{figure*}

\section{Physical causes of scatter in the STFR}
\label{sec:causes of scatter}

In this section, we discuss the vertical scatter in the stellar and gas STFRs (i.e the residuals in the STFR from a LSQ fit), and its variation with radius. For the remaining of this paper, we refer to this scatter as ${\Delta}$TF (with subscripts indicating the stellar or gas relation). This measure is calculated, for each galaxy sample, with respect to the best-fitting relation for rotators using a LSQ fit (Fig. \ref{fig:TF_rotators}), for either stars or gas, and at that respective radius. We employ the rotator sample since it produces the correlation with the least scatter at all the probed radii (see Table \ref{tab:breakdown_TF} for an overview).

To test for correlations between ${\Delta }$TF and different galaxy properties and environmental metrics, we calculate the Pearson partial linear correlation coefficient $\rho_{\rm{partial }}$. This coefficient takes values between -1 and 1 such that a larger absolute value of $\rho_{\rm{partial }}$ denotes a stronger correlation. The two extreme values ($\pm 1$) indicate a perfect linear correlation/anti-correlation, and 0 indicates no degree of linear correlation. The Pearson partial linear correlation coefficient also takes into account the covariance between the two parameters being studied and other quantities (see e.g. \citealt{varidel_resolved_2016}, \citealt{barsanti_sami_2023}).
While separations between strong and weak correlations based on the value of $\rho_{\rm{partial }}$ are arbitrary, any degree of correlation inconsistent with 0 can show statistical significance/insignificance given the sample size. To assess the significance of the partial linear correlation coefficient, we test the null hypothesis that the value of $\rho_{\rm{partial }}$ is consistent with 0 (i.e. no linear correlation). This test returns the p value (i.e. the probability) that the partial linear correlation returned by $\rho_{\rm{partial }}$ is caused by statistical chance and hence is insignificant, i.e there is no correlation between the two parameters in question for the parent population from which our sample is drawn. We reject the null hypothesis for p values below the critical level of 0.05. As such, a p value < 0.05 signifies a statistically significant correlation (whether weak or strong) which is representative of the parent population and not introduced by the selection of our sample. In all cases when $\rho_{\rm{partial }}$ is consistent with 0 within the 5-95 per cent parametric confidence interval, we expect the null hypothesis to be accepted (p value > 0.05) if the $\rho_{\rm{partial }}$ value is a true representation of the lack of correlation in the parent sample from which our data is drawn.

We analyse correlations between ${\Delta}$TF and the following galaxy properties and environmental metrics: $(v/\sigma)_{\rm{NR_{\rm{e}}}}$ (N = 1,1.3,2), log$(\Sigma_{\star}/\rm{M_{\odot}}\rm{kpc^{-2}})$, log$(\Sigma_{\rm{SFR}}/\rm{M_{\odot}}\rm{yr^{-1}kpc^{-2}})$, $n_{\rm{s}}$, T-type, $Q_{\rm{group}}$ and $\Delta \rm{PA_{\rm{st-g}}}$. For the remaining of this paper, whenever we calculate $\rho_{\rm{partial }}$ between ${\Delta}$TF and any of the above parameters, we take into account the covariance with $M_{\star}$ and all the other parameters mentioned, including integrated $(v/\sigma)$. This choice is motivated by the fact that the rotational-to-dispersional support correlates strongly with the scatter in the STFR (see Fig. \ref{fig:TF_kin}). As such, accounting for the covariance with $(v/\sigma)$ ensures that we are searching for physical causes of scatter which are not correlated with rotational-to-dispersional support themselves; in other words, we are analysing contributions to the scatter of the STFR that are not related to an increase in the velocity dispersion of the respective component. The value returned by $\rho_{\rm{partial }}$ thus reflects the degree of correlation between ${\Delta}$TF (A) and a given galaxy/environmental property (B), while excluding the contribution to this correlation due to A and B both being correlated or anticorrelated with any other metric. The partial correlation analysis is ideal for this work given that a number of parameters analysed to explain the scatter in the STFR are correlated with each other (e.g. $\Sigma_{\rm{SFR}}$ increases with T-type, such that late-type galaxies have higher SFR surface densities).

To assess the contribution of covariances with other parameters to the observed trends, we compare $\rho_{\rm{partial }}$ with the Pearson absolute linear correlation coefficient $r_{\rm{P}}$, equal to the covariance between the two parameters being studied divided by the product of their standard deviations (i.e. without accounting for covariances with other parameters). 
We also search for correlations of STFR scatter with group membership (central/isolated or satellite), and the presence of bars and rings. 

\subsection{Galaxy physical properties and environmental metrics}
\label{sec:scatter physical props}

For both the stellar and gas STFRs, the strongest correlation with ${\Delta}$TF are found, as expected, with integrated $(v/\sigma)$ within each radius for the respective tracer, such that more dispersion-supported systems are found below the stellar/gas STFRs at fixed $M_{\star}$. This correlation is highlighted on the first and third columns of Fig. \ref{fig:dBF_both}, showing a strong dependence for the stars ($\rho_{\rm{partial }}$ between 0.61 and 0.77) and a shallow correlation ($\rho_{\rm{partial }}$ between 0.36 and 0.48) for the gas at all probed radii. The same correlation is observed for the stellar and gas STFRs when only considering the common kinematic sample (pink line). The difference seen here between the two baryonic components, as discussed in Section \ref{sec:TF_kinematic_full}, can be attributed to the collisionless nature of stars that allows them to form dispersion-supported systems, resulting in the larger scatter in the stellar STFR compared to the gas one. However, different selection criteria for our stellar and gas kinematic samples are expected to play a role as well, given that galaxies with the most dispersion-dominated stellar kinematics often do not have rotation in the gas.

The trends showed in the second column of Fig. \ref{fig:dBF_both} indicate a slight anti-correlation at 1$R_{\rm{e}}$ and 1.3$R_{\rm{e}}$ ($r_{\rm{P}} = -0.14, 
-0.11$, respectively, without accounting for covariances) between stellar STFR residuals and  log$(\Sigma_{\star})$, i.e. a preference for more centrally concentrated galaxies (larger $\Sigma_{\star}$) to be found below the stellar STFR, and an approximately constant trend at 2$R_{\rm{e}}$, although with a notable reduction in number statistics below $\rm{log}(\Sigma_{\star}) < 8$. The same observation can be made for the common kinematic sample (pink line), highlighting that the individual result for the stellar kinematic sample is not driven by selection effects. However, while accounting for covariances with $M_{\star}$ and the other parameters mentioned in the section above, the result is a linear correlation with $\rho \approx 0.30$. This is the result of the initial anti-correlation being suppressed by (i) the correlation between $\Delta$TF and $(v/\sigma)$ and (ii) the anti-correlation between log($\Sigma_{\star}$) and $(v/\sigma)$, and as such holds little physical meaning. 
Furthermore, in all of these cases, the running medians for both the stellar and common kinematic samples are consistent with $\Delta $TF = 0, within the scatter ($16^{\rm{th}}$ and $84^{\rm{th}}$ percentiles).

In the case of the gas STFR, the strongest correlations outside of $(v/\sigma)$, albeit shallow (|$\rho_{\rm{partial}}$| $\leqslant\ 0.17$), are found with log$(\Sigma_{\rm{SFR}})$ and T-type. The partial correlation with T-type is a reflection of covariances with the other considered parameters, since the absolute linear correlation coefficient $r_{\rm{P}}$ is consistent with 0 in all cases ($r_{\rm{P}}$ = 0.02, 0.01 and 0.07 at 1$R_{\rm{e}}$, 1.3$R_{\rm{e}}$ and 2$R_{\rm{e}}$, without accounting for covariance). This is not the case for the trends with log$(\Sigma_{\rm{SFR}})$ which are highlighted in the rightmost column of Fig. \ref{fig:dBF_both} (albeit with $r_{\rm{P}}$ values only marginally inconsistent with 0). For both parameters however, the running medians for the gas and common kinematic samples are always consistent with $\Delta \rm{TF} =0$. We note similar dependencies for both the gas and common (pink line) kinematic samples, i.e. a slight anti-correlation between $\Delta$TF and log$(\Sigma_{\rm{SFR}})$. This result is consistent with a framework in which star formation feedback or gravitational instabilities associated with star forming regions have the (relatively small) statistical effect of perturbing gas kinematics, causing offsets from the STFR (which are not correlated with an increase in gas $v/\sigma$). This offset from the STFR potentially caused by star formation feedback or gravitational instabilities is less pronounced at 1$R_{\rm{e}}$ ($\rho_{\rm{partial}} = -0.08$), and somewhat higher at 2$R_{\rm{e}}$ ($\rho_{\rm{partial}} = -0.18$), potentially indicating a scenario in which there is more star formation occurring in the outer edges (2$R_{\rm{e}}$) compared to the inner parts (1$R_{\rm{e}}$) in the galaxies in our gas kinematic sample. We note, however, the relative shallowness of this correlation, with running medians that are consistent with $\Delta$TF = 0 at all radii in the case of log$(\Sigma_{\rm{SFR}})$.

\begin{figure*}
	\centering
	\includegraphics[width=\linewidth]{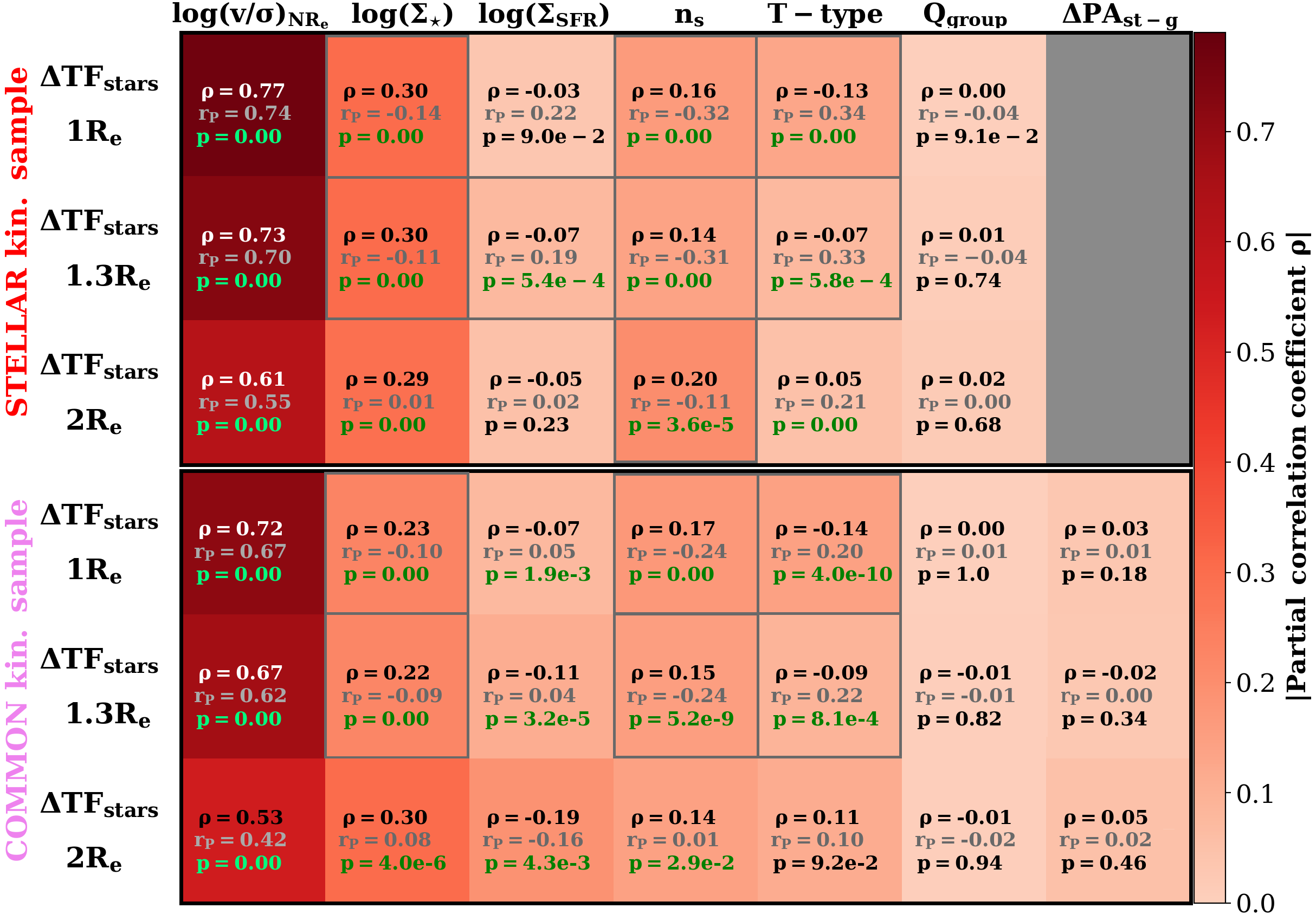}
    \caption{Grid plot showing the correlation between \textcolor{red}{\textbf{stellar}} STFR vertical scatter ($\rm{\Delta} TF_{\rm{stars}}$) and the parameters shown on the top axis. Results are shown for the stellar STFR at 1$R_{\rm{e}}$, 1.3$R_{\rm{e}}$ and 2$R_{\rm{e}}$, as displayed on the left of each row. The \textbf{top} grid shows the correlation for the \textit{stellar kinematic sample}, while the bottom displays the same results for the \textit{common kinematic sample}. Each cell is color-coded by the partial linear correlation coefficient $\rho_{\rm{partial }}$, which takes into account the covariance with stellar mass and all the other parameters evaluated. The partial ($\rho_{\rm{partial }}$) and absolute ($r_{\rm{P}}$) linear correlation coefficients  are displayed on each cell. The grey squares highlight cells where there is a sign difference between the partial and absolute linear correlation coefficients, and both are not consistent with 0, within uncertainties. The p value displayed on each cell tests the null hypothesis that $\rho_{\rm{partial }}$ is consistent with 0. Values below 0.05 are considered to reflect a statistically significant correlation at the 5 per cent confidence level, and are shown in green. In all cases where the $\rho_{\rm{partial }}$ value is consistent with 0 within the uncertainties (5-95 per cent confidence interval), we obtain a p value > 0.05 (shown in black), signifying that the lack of correlation is significant. The 5-95 per cent confidence intervals for $\rho_{\rm{partial }}$ are between $\pm0.02$ and $\pm 0.06$ at 1$R_{\rm{e}}$ and 1.3$R_{\rm{e}}$, and between $\pm0.05$ and $\pm 0.10$ at 2$R_{\rm{e}}$. }
    \label{fig:grid_st}
\end{figure*}

We also keep track of underlying correlations with stellar mass, visualised by the color coding of Fig. \ref{fig:dBF_both}. We note no correlations of STFRs residuals with stellar mass, a result also confirmed by computing the Durbin-Watson statistic which accounts for the degree of correlation between consecutive STFR residuals (\citealt{durbin_testing_1950}). For the entirety of our kinematic and rotator sub-samples, the values of the Durbin-Watson statistic are in the range 1.75-2.14 and 1.51-2.05 respectively, which indicates randomly distributed residuals that follow a normal distribution. As such, our study finds no statistical evidence of a bending of the STFR at high stellar masses, for either stellar or gas rotational velocity. Previous studies have only reported very marginal evidence of such a bending in the STFR (see e.g. \citealt{boubel_preprint_2023} and references therein). 

\begin{figure*}
	\centering
	\includegraphics[width=\linewidth]{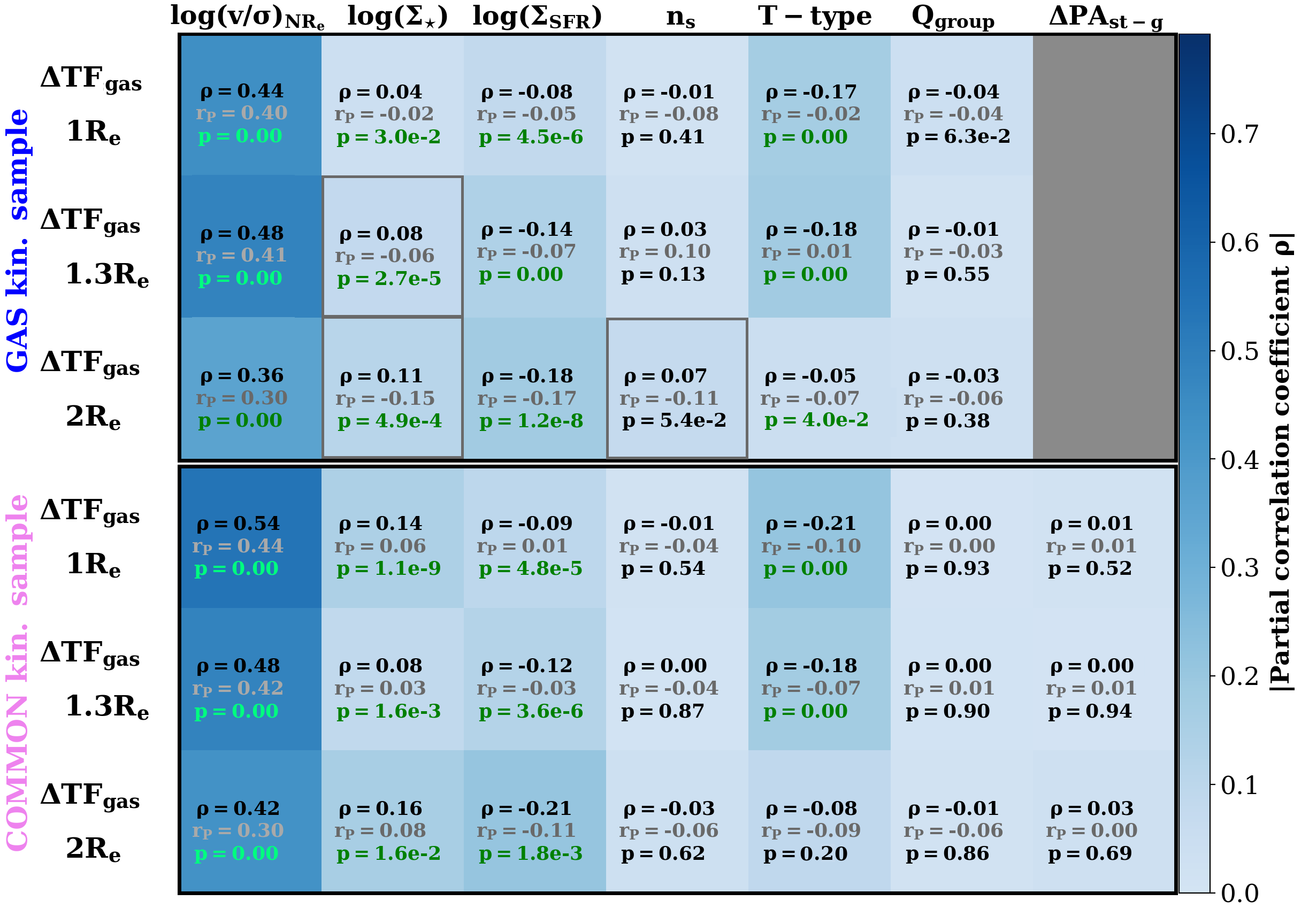}
    \caption{The same as Fig. \ref{fig:grid_st}, for the \textcolor{blue}{\textbf{gas}} STFR computed for the gas kinematic sample (\textbf{top}) and common kinematic sample (\textbf{bottom}).}
    \label{fig:grid_g}
\end{figure*}

The linear correlation coefficients for the entire suite of parameters analysed in this paper are shown in Fig. \ref{fig:grid_st} and Fig. \ref{fig:grid_g} for the stellar and gas STFRs, respectively. To keep track of sample biases, we show the results for the common kinematic sample in the bottom part of each figure as well. The cells highlighted in grey reflect cases where the partial and absolute linear correlation coefficients have different signs (and both coefficients are not consistent with 0). This scenario suggests that the absolute correlation between two parameters ($r_{\rm{P}}$) is smaller than the sum of their covariances with the other metrics considered in the partial correlation analysis. As such, the value of $\rho_{\rm{partial}}$ in these cases does not reflect the true (uncorrected for covariances) correlation between the parameters of interest, but is rather only a reflection of the strong covariances between the two parameters of interest and the tertiary metrics that are being corrected for.

The p values on each cell test the null hypothesis that $\rho_{\rm{partial}}$ is consistent with 0, and are shown in green if p < 0.05 and the hypothesis is rejected. In all cases where p > 0.05, the value of $\rho_{\rm{partial}}$ is consistent with 0 within uncertainties, reflecting a lack of statistically significant correlation.

In the case of the stellar STFR (Fig. \ref{fig:grid_st}), for both our stellar (top) and common (bottom) kinematic samples, correlations with log$(\Sigma_{{\rm{SFR}}})$ are very shallow (|$\rho|\leqslant0.19$) at all radii. These correlations are only statistically significant (p-value < 0.05, with $\rho_{\rm{partial}}$ and $r_{\rm{P}}$ having the same sign) for the common kinematic sample, at all radii. In the case of $Q_{\rm{group}}$, $\rho_{\rm{partial}}$ is consistent with 0 at all radii and for both the stellar kinematic and common samples. For the common kinematic sample, we can assess correlations with the kinematic misalignment angle between stars and gas $\Delta \rm{PA}_{\rm{st-g}}$, finding no correlations (|$\rho|\leqslant0.05$, consistent with 0) between this metric and STFR residuals at any radius. Our proxies for optical morphology ($\Sigma_{\star}$, T-type and $n_{\rm{s}}$) show no statistically significant correlations with STFR residuals for all cases at 1$R_{\rm{e}}$ and 1.3$R_{\rm{e}}$, as shown by the sign difference between the absolute and partial coefficients. While this is not the case at 2$R_{\rm{e}}$ for $\Sigma_{\star}$, T-type and $n_{\rm{S}}$ (except for $n_{\rm{S}}$ when using the stellar kinematic sample, and T-type when using the common kinematic sample), the correlations recovered are relatively shallow ($\rho \leqslant 0.30$). 
These results for the stellar STFR suggest that none of the parameters probed in this work other than $v/\sigma$ ratio (global SFR surface density, tidal interaction strength or stellar-gas kinematic misalignments indicating a recent accretion event; see e.g. \citealt{ristea_sami_2022}) correlate strongly with STFR residuals, while also not being related with rotational-to-dispersional support. Once the $v/\sigma$ dependence is taken into account, the remaining scatter does not encode any strong physical meaning (i.e a strong correlation with galaxy properties or environment) in the inner parts (1$R_{\rm{e}}$ and 1.3$R_{\rm{e}}$), as far as the parameters analysed in this work probe. At 2$R_{\rm{e}}$, there are  only shallow correlations (albeit statistically significant) between STFR scatter and optical morphology (log($\Sigma_{\star}$), T-type and $n_{\rm{s}}$).

The full set of gas STFRs residuals correlations are shown in Fig. \ref{fig:grid_g}. There are only shallow correlations with optical morphology and SFR surface density (|$\rho_{\rm{partial}}<0.21$|), albeit statistically significant except for $n_{\rm{S}}$. Correlations with group tidal strength or stellar-gas kinematic misalignments for both the gas and common kinematic samples are consistent with 0 in all cases (p values < 0.05). As such, the above considerations for the stellar STFR are valid in the case of the gas as well: once the correlation of gas STFR scatter with turbulence parameterised through $v/\sigma$ is taken into account, the remaining scatter does not appear to encode any physical meaning, i.e. the intrinsic scatter in the gas STFR appears to be largely explained by the variation in $v/\sigma$.

While the results in Fig. \ref{fig:grid_st} and Fig. \ref{fig:grid_g} indicate no linear correlation between stellar-gas kinematic misalignments ($\rm{\Delta PA_{st-g}}$) and the scatter with respect to the stellar and gas STFRs, there is no physical evidence suggesting that a larger misalignment angle would produce a larger scatter from the relation. 
We therefore explore the possibility that kinematically misaligned galaxies are scattered preferentially below or above the stellar and gas STFRs in Appendix \ref{sec:Appendix_A}. We find that the majority of galaxies in our common kinematic sample are found below both the stellar and gas STFRs, and have typically lower $v/\sigma$ rations than aligned objects.

We note that our trends at different radii in Fig. \ref{fig:grid_st} and \ref{fig:grid_g} are computed for different samples of decreasing size towards larger probed radii. To keep track of the potential effect of the sample biases presented in Section \ref{sec:final_sample}, we re-compute the correlation coefficients for the STFRs at 1$R_{\rm{e}}$ and 1.3$R_{\rm{e}}$ only for the sample of galaxies with both stellar and gas kinematics reaching $2R_{\rm{e}}$. In the case of correlations with integrated $v/\sigma$, as expected, we see a reduction in $\rho_{\rm{partial }}$ by 0.16 and 0.17 for stellar kinematics at 1$R_{\rm{e}}$ and 1.3$R_{\rm{e}}$. The reduction for gas kinematics at the same radii is by 0.31 and 0.28, respectively. This significant change is indicative of the lower dynamical range in $v/\sigma$ for the common kinematic sample reaching 2$R_{\rm{e}}$ (compared to sub-samples reaching 1$R_{\rm{e}}$ and 1.3$R_{\rm{e}}$), as discussed in Section \ref{sec:TF_kinematic_full}. For all the other parameters in Fig. \ref{fig:grid_st} and Fig. \ref{fig:grid_g}, we find relatively small changes in $\rho_{\rm{partial }}$ (within 0.18) for both stars and gas at all radii, and in all cases resulting in correlations of a lower magnitude than presented on the figures. The statistical significance of each correlation as determined by the p values displayed in Fig. \ref{fig:grid_st} and Fig. \ref{fig:grid_g} is maintained. 

In summary, our results suggest that, once the dispersion support in stars and gas is taken into account, the scatter in the stellar and gas STFRs shows no strong correlations with either: optical morphology, SFR surface density, group tidal interaction strength or kinematic misalignment angle (indicating a recent accretion event). These findings are consistent with a scenario in which the increase in stellar and/or gas velocity dispersion support from either external processes (e.g. galaxy mergers, gas accretion) or internal ones (e.g. feedback from star formation or active galactic nuclei) is the dominant and fundamental cause of scatter in the stellar and gas STFRs.

\begin{figure}
	\centering
	\includegraphics[width=\columnwidth]{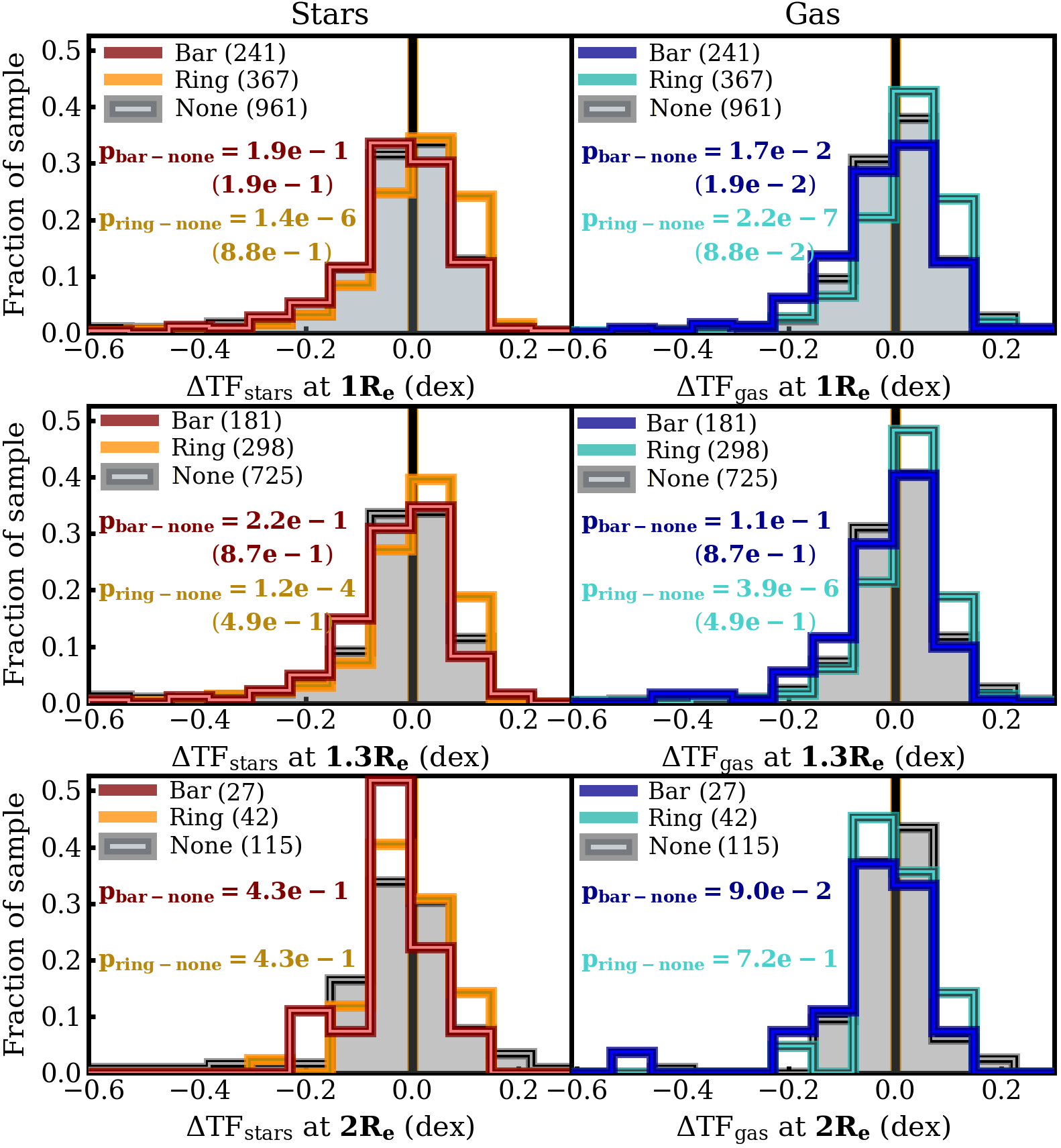}
    \caption{The distribution of \textcolor{red} {\textbf{stellar}} (left) and \textcolor{blue} {\textbf{gas}} (right) STFR residuals at fixed stellar mass ($\Delta \rm{TF}_{\rm{stars/gas}}$) for the relations computed at 1$R_{\rm{e}}$, 1.3$R_{\rm{e}}$ and 2$R_{\rm{e}}$ (top, middle and bottom, respectively), split according to the presence or absence of bars/rings. The number of galaxies in each morphological category is shown in the legend, in brackets. The distributions correspond to the common kinematic sample (see Section \ref{sec:final_sample}). Distributions for galaxies with bars/rings are shown in red/orange (stellar STFR) and blue/cyan (gas STFR), while distributions for objects with no identified morphological features are shown in grey. The p values displayed on each panel are the result of a comparison (using a KS test) of the distributions of galaxies with no features and those with bars/rings, as indicated in the subscript. The values in brackets refer to the same results when considering only the common kinematic sample reaching 2$R_{\rm{e}}$ (i.e. the sample of galaxies on the last row).}
    \label{fig:morpho_hist}
\end{figure}

\subsection{Bars and rings}
\label{sec:bars_rings}

We also test whether the presence of bars and/or rings in galaxies correlates with their position with respect to the STFRs. To ensure a fair comparison between stellar and gas kinematics, we only analyse our common kinematic sample (galaxies with both stellar and gas kinematics; see Section \ref{sec:final_sample}). We split this sample into galaxies with bars, rings or no morphological features (identified as described in Section \ref{sec:data_and_methods}). The distributions of stellar/gas STFRs residuals for each of these sub-samples are shown in Fig. \ref{fig:morpho_hist}, with the number of galaxies in each sub-sample displayed in the legend. We compare the distributions of $\Delta$TF for galaxies with bars/rings and those with no identified morphological feature using a KS test, with results shown on each panel of Fig. \ref{fig:morpho_hist}.

At 1$R_{\rm{e}}$ and 1.3$R_{\rm{e}}$, we find statistical dissimilarities between the STFR residual distributions of ring and non-feature galaxies (p values between $2.2\times 10^{-7}$ and $1.2\times 10^{-4}$), for both the stellar and gas STFRs (top and middle rows of Fig. \ref{fig:morpho_hist}). In the case of bars, we only note a borderline statistical difference at the 5 per cent level for the gas STFR at 1$R_{\rm{e}}$ (p value = 0.017).

While the above results are statistically significant, we must first acknowledge the limitations of the visual classification scheme employed by Galaxy Zoo, especially in terms of separating galaxies which only have a ring/bar from those with both components, as well as the uncertainties in defining effective radii in barred/ring galaxies, given their light profiles. Due to the bar selection threshold applied for our bar and ring selection (see \citealt{fraser-mckelvie_sdss-iv_2020}), our results are expected to include preferentially strong bars/rings, and might fail to identify weak features.

Furthermore the comparison between results at different radii in Fig. \ref{fig:morpho_hist} is impeded by the fact that panels on the top, middle and rows reflect results for different galaxy samples. We perform a check by only considering the galaxies in the common kinematic sample with velocity measurements reaching 2$R_{\rm{e}}$, and re-compare the relevant distributions in Fig. \ref{fig:morpho_hist}
at 1$R_{\rm{e}}$ and 1.3$R_{\rm{e}}$ (with p values displayed in brackets). This selection eliminates the statistical differences noted between the STFR residuals distributions of ring galaxies and those with no features, for both the stellar and gas relations (top and middle of Fig. \ref{fig:morpho_hist}), with the p values being in the range 0.09-0.88. This change is the result of the biases introduced when only selecting galaxies with stellar and gas kinematics reaching 2$R_{\rm{e}}$. As such, the dissimilarities mentioned above do not hold for the common kinematic sample reaching 2$R_{\rm{e}}$, which largely includes massive main-sequence galaxies. The differences noted for the entire common kinematic sample at 1$R_{\rm{e}}$ and 1.3$R_{\rm{e}}$ are thus driven by low stellar mass galaxies in the sub-samples reaching these radii (see Section \ref{sec:final_sample} and Fig. \ref{fig:sample_plot}), for which ring and bar classifications are expected to be more uncertain than in the massive galaxies with kinematics reaching 2$R_{\rm{e}}$.

If physically meaningful given the mentioned uncertainties, the differences in $\Delta \rm{TF}$ distributions between ringed and non-feature galaxies at 1$R_{\rm{e}}$ and 1.3$R_{\rm{e}}$ for our entire common kinematic sample are manifested in the form of ring galaxies being preferentially found above the stellar and gas STFRs defined for rotators. Such differences could potentially be attributed to a redistribution of angular momentum associated with the ring formation process, which largely occurs in cold rotating discs with a strong bulge or thick disc component. We note however that no such discrepancies are found in the case of barred galaxies, in accord with the findings of \cite{courteau_tully-fisher_2003}, and suggesting a similar dynamical behaviour of barred and no-feature galaxies.
While it is believed that the formation pathways of bars and rings are interconnected, the intricacies of this connection are not entirely understood, with up to a third of galaxies hosting inner rings not having bars (\citealt{diaz-garcia_inner_2019}). Given the inherent uncertainty of our feature classification, we do not make a separation between galaxies that \textit{only} exhibit a bar/ring.

Our results indicate little contribution from processes forming strong bars and rings on the position of galaxies withe respect to the stellar and gas STFRs. When the entire common kinematic sample is considered, tentative differences are noted in the form of ringed galaxies being found preferentially more above the STFRs for stars and gas compared to objects with no morphological features, albeit with notable uncertainties given our bar/ring classification scheme, as discussed.

\subsection{Group membership}
\label{sec:group membership}

\begin{figure}
	\centering
	\includegraphics[width=\columnwidth]{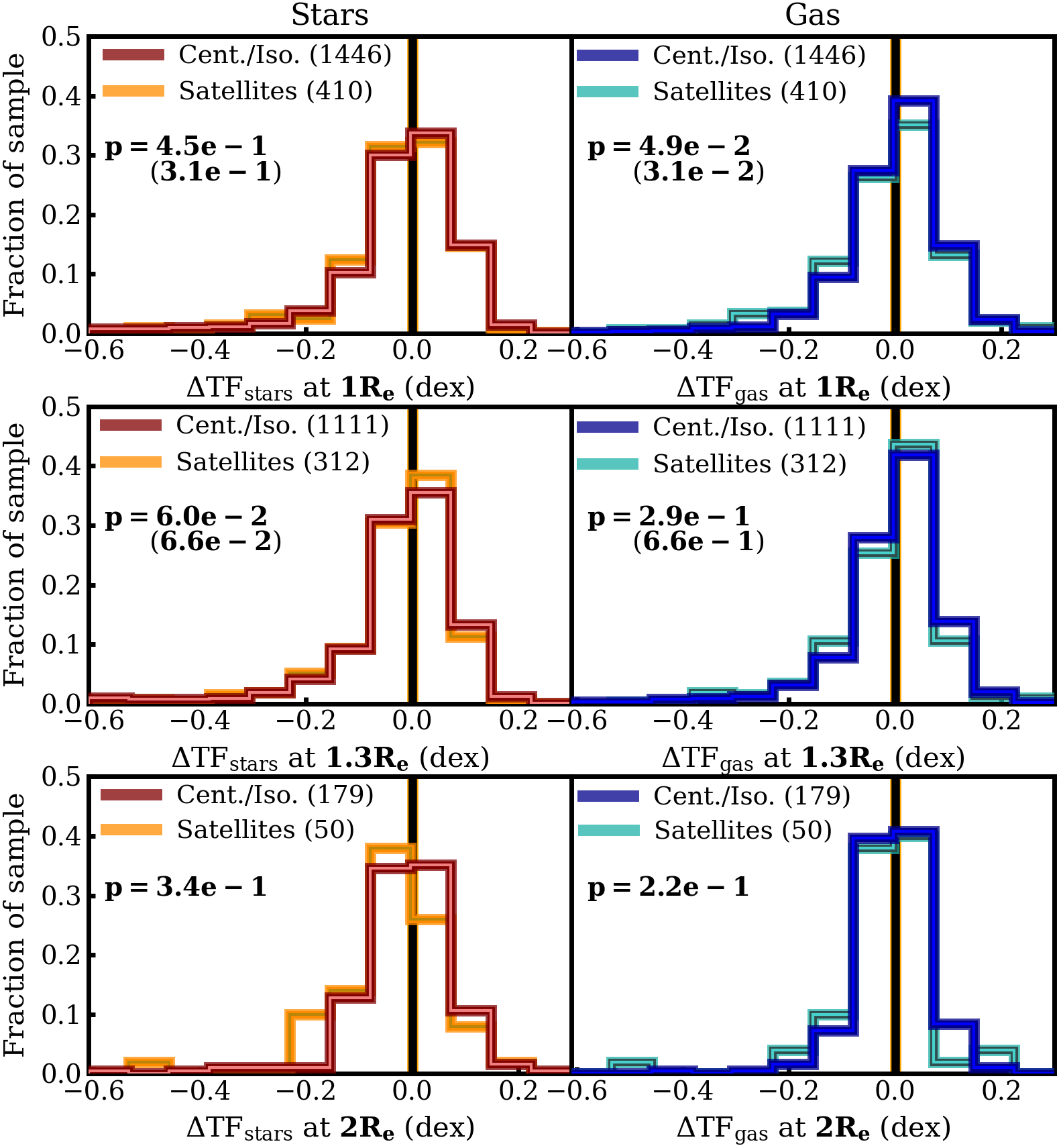}
    \caption{The distribution of \textcolor{red} {\textbf{stellar}} (left) and \textcolor{blue} {\textbf{gas}} (right) STFR residuals ($\Delta \rm{TF}_{\rm{stars/gas}}$) for the relations computed at 1$R_{\rm{e}}$, 1.3$R_{\rm{e}}$ and 2$R_{\rm{e}}$ (top, middle and bottom, respectively), split according to the group membership of galaxies (centrals/isolated and satellites). The number of galaxies in each environmental category is shown in the legend, in brackets. The distributions correspond to the common kinematic sample (see Section \ref{sec:final_sample}). Distributions for centrals or isolated galaxies are shown in red (stellar STFR) and blue (gas STFR), while distributions for group satellites are shown in orange (stellar STFR) and cyan (gas STFR). The p values displayed on each panel are the result of a comparison (using a KS test) of the distributions of centrals/isolated galaxies and those that are group satellites. The values in brackets refer to the same results when considering only the common kinematic sample reaching 2$R_{\rm{e}}$ (i.e. the sample of galaxies on the last row). }
    \label{fig:env_hist}
\end{figure}

We test the possibility of the galactic environment having an effect on the position of galaxies with respect to the stellar and/or gas STFRs. We again select our common kinematic sample for this purpose (Section \ref{sec:final_sample}) and split it into centrals or isolated galaxies and group satellites. We show the distributions of stellar and gas STFR residuals at 1$R_{\rm{e}}$, 1.3$R_{\rm{e}}$ and 2$R_{\rm{e}}$ for the central/isolated and satellite sub-samples in Fig. \ref{fig:env_hist}, with the number of galaxies in each sub-sample displayed in the legend.

In all cases of the stellar STFR (Fig. \ref{fig:env_hist}, left side), the distributions of $\Delta$TF for centrals/isolated and satellites show statistical similarities at the 5 per cent level (p value between 0.06 and 0.45). The same result is noted for the gas STFR (Fig. \ref{fig:env_hist}, right side), with the exception of the case at 1$R_{\rm{e}}$ where only marginal statistical differences are found (p value = 0.049).

We acknowledge again that the results at different radii in Fig. \ref{fig:env_hist} are representative of different galaxy sub-samples of the common kinematic sample, and thus not directly comparable. We only consider the galaxies with kinematics reaching 2$R_{\rm{e}}$ and re-compare the relevant distributions for the STFRs at 1$R_{\rm{e}}$ and 1.3$R_{\rm{e}}$. The results of this comparison are displayed in brackets in Fig. \ref{fig:env_hist}. We note no changes in the statistical significance of our results when only considering the common kinematic sample reaching 2$R_{\rm{e}}$, compared to the full common kinematic sample.

The similarities in stellar and gas $\Delta \rm{TF}$ distributions at all radii between centrals/isolated and satellite galaxies suggest a negligible statistical effect of group environment processes on the position of galaxies with respect to the STFR, with the mention that the largest group probed by the MaNGA Survey is only numbering 623 galaxies.

\section{Summary and conclusions}
\label{sec:conclusion}

We have performed an analysis of the STFR for stellar and gas rotation at different radii in the nearby Universe, without any selection based on optical or kinematic morphology. We have shown how the stellar and gas STFRs change if we instead only select rotationally dominated objects, as is the canonical method of STFR studies. We have discussed the implications of sample matching on the radial variation of the STFR, and on the differences between the stellar and gas relations. We have also performed an analysis of the scatter in the stellar and gas STFRs and examined its correlations with various galaxy properties and environmental metrics.

Our main conclusions are as follows:
\begin{itemize}
    \item \textbf{When the full samples (kinematic/rotator) are considered} (Section \ref{sec:TF_kinematic_full}), the stellar and gas STFRs are becoming shallower between 1$R_{\rm{e}}$ and 2$R_{\rm{e}}$ (Fig. \ref{fig:TF_kin}) indicating the existence of different, independent relations up to 2$R_{\rm{e}}$ for both components. This set of relations is the result of different galaxy samples having different rotation curve shapes that vary depending on stellar mass (\citealt{catinella_template_2006}; \citealt{yoon_rotation_2021}). The stellar STFRs best fit parameters are consistent within errors with those of the gas STFR at all probed radii. In the case of our rotator samples, the radial variation is the same as for the full kinematic samples, i.e the relations become slightly shallower as the radius used to probe velocity increases, albeit with the slopes being consistent within uncertainties at all radii, for both baryonic components. This trend is highlighted in Fig. \ref{fig:TF_rotators} and indicates that our rotator sample selection includes galaxies with little variation in stellar and gas dynamical state between 1$R_{\rm{e}}$ and 2$R_{\rm{e}}$. The intrinsic scatter in the stellar STFR for the entire kinematic sample is slightly larger than in the gas at 1$R_{\rm{e}}$ and 1.3$R_{\rm{e}}$, while the two are consistent at 2$R_{\rm{e}}$. The intrinsic scatters in the stellar and gas STFR are also consistent when only considering the rotator sample (Table \ref{tab:breakdown_TF}). Overall, these findings confirm results previously reported by \cite{yegorova_radial_2007}, and extend them up to 2$R_{\rm{e}}$.

    \item When comparing the stellar and gas STFRs for rotators with previous literature computations of the STFR (Fig. \ref{fig:TF_comparison_samples}), we recover a good agreement (<0.02 dex in log(velocity) at any given stellar mass) in the case of stellar rotation (\citealt{brownson_what_2022}). When considering relations that trace gas rotations from either H$\rm{\alpha}$ (\citealt{bloom_sami_2017}, \citealt{catinella_sami-h_2023}, \citealt{arora_manga_2023}) or \ion{H}{i} (\citealt{catinella_sami-h_2023}), we identify a slightly larger spread (up to $\sim$ 0.19 dex in log(velocity) for our range of $M_{\star}$), especially at log($M_{\star}/\rm{M_{\odot}}$) < 10. This discrepancy can be potentially attributed to either the use of galaxy samples representative of different populations, different radii used for velocity estimation or different stellar mass computation methods (e.g using \textit{g-i} magnitudes as opposed to full SED fitting, which are expected to diverge at lower stellar masses).

    \item \textbf{When only considering galaxies with stellar and gas kinematics reaching 2$R_{\rm{e}}$} (Section \ref{sec:TF_kinematics_matched}), we recover steeper stellar and gas STFR than in the unmatched case (Fig. \ref{fig:TF_comparison_sample_match}), although differences in the gas relation are smaller than in the stars between the matched and unmatched cases. We also obtain the same radial trends of the STFRs as in the case where the full kinematic samples are considered (i.e. the relations for both components become shallower as the radius used to probe the velocity increases). We report a shallower gas STFR at 1$R_{\rm{e}}$ and 1.3$R_{\rm{e}}$ compared to the stellar one, while in the outer edges (2$R_{\rm{e}}$), the relations for the two components are consistent within uncertainties. This finding indicates that, in the nearby Universe, the stars and gas are following different states of dynamical equilibrium in the inner parts ($\leqslant$1.3$R_{\rm{e}}$) due to larger contributions from random motions for the stellar component. The two baryonic components are dynamically coupled in the outer edges (2$R_{\rm{e}}$). The intrinsic scatter in the stellar STFRs is consistent with the one in the gas within 0.01 dex, when the same galaxy sample is considered. This similarity indicates that the previously reported larger intrinsic scatter at 1$R_{\rm{e}}$ and 1.3$R_{\rm{e}}$ in the stellar STFR when the full kinematic samples were used is largely due to different sample selection effects. When analysing the stellar and gas relations for our rotator samples, there is no significant radial variation, while the relations for stars and gas are consistent at each radius. This result suggests that when only rotationally dominated galaxies in the nearby Universe are considered, the two baryonic components are statistically found the same state of dynamical equilibrium up to 2$R_{\rm{e}}$.

    \item No strong correlations with STFR vertical offset are noted for any of the probed physical parameters and environmental metrics, for either stars or gas (see Fig. \ref{fig:grid_st} and Fig. \ref{fig:grid_g}). This result suggests that, once the dispersion support in galaxies is corrected for, the remaining scatter shows only shallow correlations with optical morphology and SFR surface density, and no correlation with environmental tidal metrics or signatures of recent gas accretion. Ring galaxies are only marginally found above the stellar and gas STFR compared to objects with no features (Fig. \ref{fig:morpho_hist}, only valid when the full common kinematic sample is considered), a result not recovered in the case of barred galaxies. These findings indicate a negligible contribution from the processes forming strong bars and rings to driving the scatter in the STFR. Furthermore, the distributions of STFR residuals for central/isolated and satellite galaxies are statistically similar at all probed radii (Fig. \ref{fig:env_hist}). Group processes thus do not appear to have a statistically significant effect on the position of galaxies with respect to the stellar and gas STFRs.
\end{itemize}

Our results have reinforced the importance of considering biases due to sample selection or the choice of a specific kinematic tracer (stars or gas) when undertaking studies of the STFR. We report evidence for the existence of different STFRs for stellar and gas rotation at different radii, and for different kinematic morphologies, indicative of statistically different states of dynamical equilibrium for the respective baryonic tracers. Finally, our findings suggest a scenario in which the increase in the velocity dispersion support of stars and/or gas (from either galaxy mergers, gas accretion, feedback from star formation or active galactic nuclei) is the dominant and fundamental cause of scatter in the stellar and gas STFRs.
The compendium of relations presented in Table \ref{tab:breakdown_TF} and Table \ref{tab:appendix_table} provides a benchmark, bias-informed calibration tool for simulations of galaxy evolution, and a comparison point for high-redshift studies of the STFR. 

\section*{Acknowledgements}

We thank the anonymous reviewer for a constructive report which significantly improved the quality and impact of this work.
\textbf{AR} acknowledges that this research was carried out while the author was in receipt of a Scholarship for International Research Fees (SIRF) and an International Living Allowance Scholarship (Ad Hoc Postgraduate Scholarship) at The University of Western Australia. Parts of this research were conducted by the Australian Research Council Centre of Excellence for All Sky Astrophysics in 3 Dimensions (ASTRO 3D), through project number CE170100013.
\textbf{LC} acknowledges support from the Australian Research Council Discovery Project and Future Fellowship funding schemes (DP210100337, FT180100066). \textbf{JvdS} acknowledges support of an Australian Research Council Discovery Early Career Research Award (project number DE200100461) funded by the Australian Government.

This research has made use of the following Python packages: \textsc{Seaborn} \citep{waskom_seaborn_2021}, \textsc{Matplotlib} \citep{hunter_matplotlib_2007}, \textsc{SciPy} \citep{virtanen_scipy_2020}, \textsc{NumPy} \citep{harris_array_2020}, \textsc{Astropy} \citep{robitaille_astropy_2013} and \textsc{Pingouin} \citep{vallat_pingouin_2018}.

\section*{Data Availability}

The MaNGA data cubes and value-added products used in this paper are available at \url{https://www.sdss4.org/dr17/manga/}. Computed stellar and gas rotational velocities and $v/\sigma$ values at/within 1$R_{\rm{e}}$, 1.3$R_{\rm{e}}$ and 2$R_{\rm{e}}$ for the samples used in this work are outlined in the Supplementary Information (MaNGA kinematic catalogue section, available online only). Further data can be provided upon request to the author.

 \section*{Author contribution statement}

This project was devised by \textbf{AR}, \textbf{LC} and \textbf{AFM}. \textbf{AR} performed the analysis and drafted the paper. \textbf{AFM} performed the measurements of $v/\sigma$ values used in this work. \textbf{BC}, \textbf{JvdS}, \textbf{SMC} and \textbf{AMS} discussed the results and commented on the manuscript.



\bibliographystyle{mnras}
\bibliography{Reference_Paper.bib} 




\appendix

\section{Orthogonal regression fit to the STFR}
\label{sec:Appendix_Ortho}

We re-fit each data set outlined in Table \ref{tab:breakdown_TF} using orthogonal linear regression with intrinsic scatter implemented by the \texttt{HyperFit} package \citep{robotham_hyper-fit_2015}. In doing so, we assume a constant uncertainty in stellar mass based on the maximum uncertainty in stellar mass for our galaxy sample (0.08 dex), computed as described in \cite{salim_lessigreatergalexlessigreater_2016}. We present the results of these fits in Table \ref{tab:appendix_table}. 

A comparison between the results of the orthogonal fit (Table \ref{tab:appendix_table}) and those from the LSQ fit (Table \ref{tab:breakdown_TF}) reveals a general agreement between the two methods. In all cases, the slopes from the two fitting methods agree within uncertainties. Disagreements in the intercepts larger than the error are only found for the stellar STFR at 2$R_{\rm{e}}$ in the cases of the full kinematic and rotator samples, and for the gas STFR at 1$R_{\rm{e}}$ in the case of the common kinematic sample.

Finally, we find that the same trends with scatter reported in Section \ref{sec:TF_kinematic_full} and Section \ref{sec:TF_kinematics_matched} are also valid for the orthogonal intrinsic scatter $s_{\perp}$. The intrinsic orthogonal scatter decreases as the radius used to probe velocity increases. $s_{\perp}$ is slightly higher in the stellar STFR compared to the gas one at the same radius for the full kinematic sample ($\Delta s_{\perp} \sim 0.02-0.03$ dex, albeit within uncertainties), and comparable (within 0.01 dex) for the common kinematic and rotator samples.  

\renewcommand{\arraystretch}{1.0}
\begin{table*}
\centering
\caption{Compilation of the stellar and gas STFR best-fit parameters from an orthogonal regression fit, for all the galaxy samples used in this work. We show results for our full stellar, gas and common kinematic samples as well as for the selected rotators, at 1$R_{\rm{e}}$, 1.3$R_{\rm{e}}$ and 2$R_{\rm{e}}$. We also present the best-fit relations at 1$R_{\rm{e}}$ and 1.3$R_{\rm{e}}$ only for the common kinematic and rotator samples with stellar and gas kinematics reaching 2$R_{\rm{e}}$. The number of galaxies in each stellar and gas sub-sample at the respective radius is highlighted under the \textcolor{red}{$\rm{N_{gal,st}}$} and \textcolor{blue}{$\rm{N_{gal,g}}$} columns, respectively. The values in square brackets show the uncertainty in the last decimal place of each best-fit value. $s_{\perp}$ is the best-fit orthogonal intrinsic scatter, expressed in dex. The intercept (b) values are expressed in dex, while the slopes (a) are dimensionless.  The colored cells correspond to the relations shown in Fig. \ref{fig:TF_kin} (\textcolor{red}{red} and \textcolor{blue}{blue}) and Fig. \ref{fig:TF_rotators} (\textcolor{chromeyellow}{yellow} and \textcolor{caribbeangreen}{green}). }
\setlength\tabcolsep{1.9pt}
\label{tab:appendix_table}
\begin{tabular*}{\linewidth}{@{\extracolsep{\fill}} c c|ccc|ccc} 
\multicolumn{1}{c}{\multirow{2}{*}{   Sample   }}                                     & \multirow{2}{*}{   Radius $(\times R_{\rm{e}})$
   | \textcolor{red}{$\rm{N_{gal,st}}$} | \textcolor{blue}{$\rm{N_{gal,g}}$}   }  & \multicolumn{3}{c|}{Stars}                                                                                                                            & \multicolumn{3}{c}{Gas}                                                                                                               \\          
\multicolumn{1}{c}{}                                                            &                               & slope (a)                                               & intercept (b)                                              & $s_{\perp}$                       &      slope (a)                                           & intercept (b)                                          & $s_{\perp}$                            \\ 

\hline
\hline
\multirow{6.6}{*}[3em]{\begin{tabular}[c]{@{}c@{}}\\  Kinematic\\(ALL) 
 \end{tabular}}       & $\ \ \ \ \ \ \ \ $ 1   $ \ \ \ \ \ \ \ \ \ $   | $\ $ \textcolor{red}{2683}  | \textcolor{blue}{3430}               & {\cellcolor[rgb]{1,0.373,0.373}}\begin{tabular}{c}\ \ \ \ \ \    0.284[6] \ \ \ \ \ \   \end{tabular} & {\cellcolor[rgb]{1,0.373,0.373}}\begin{tabular}{c} \ \ \ \ \ \ -0.86[7] \ \ \ \ \ \ \end{tabular} & {\cellcolor[rgb]{1,0.373,0.373}}   \ \  \ \ \ 0.10[3] \ \ \ \ \ \ & {\cellcolor[rgb]{0,0.631,1}} \begin{tabular}{c}\ \ \ \ \ \ 0.283[2] \ \ \ \ \ \ \end{tabular} & {\cellcolor[rgb]{0,0.631,1}} \begin{tabular}{c} \ \ \ \ \  -0.76[3] \ \ \ \ \ \ \end{tabular} & {\cellcolor[rgb]{0,0.631,1}}\ \ \ \ \ \  0.08[2] \ \ \ \ \ \    \\
                                                                                & $\ \ \ \ \ \ \ \ $ 1.3   $ \ \ \ \ \ \  $   | $\ $ \textcolor{red}{2211}  | \textcolor{blue}{3047}                              & {\cellcolor[rgb]{1,0.373,0.373}}\begin{tabular}{c} 0.266[6]\end{tabular}               & {\cellcolor[rgb]{1,0.373,0.373}}\begin{tabular}{c} -0.63[6] \end{tabular}                & {\cellcolor[rgb]{1,0.373,0.373}}0.10[3]        &  {\cellcolor[rgb]{0,0.631,1}} \begin{tabular}{c} 0.267[3]\end{tabular}                 & {\cellcolor[rgb]{0,0.631,1}}\begin{tabular}{c} -0.58[3]
                                                \end{tabular}                & {\cellcolor[rgb]{0,0.631,1}}0.07[2]             \\
                                                                               & $\ \ \ \ \ \ \ \ $ 2   $ \ \ \ \ \ \ \ \ \ $   | $\ $ \textcolor{red}{530} $\ $  | \textcolor{blue}{1019}                              & {\cellcolor[rgb]{1,0.373,0.373}}\begin{tabular}{c} 0.25[1] \end{tabular}                & {\cellcolor[rgb]{1,0.373,0.373}}\begin{tabular}{c} -0.43[1]
                                                                                \end{tabular}               & {\cellcolor[rgb]{1,0.373,0.373}}0.09[4]          & {\cellcolor[rgb]{0,0.631,1}}\begin{tabular}{c} \ 0.241[4]
                                                \end{tabular}                & {\cellcolor[rgb]{0,0.631,1}}\begin{tabular}{c} -0.28[4] \end{tabular}               & {\cellcolor[rgb]{0,0.631,1}}0.06[3]     \\

\midrule
                                                            
\multirow{-1.66}{*}{\begin{tabular}[c]{@{}c@{}}\\  Common \\ (stars $\&$ gas) \end{tabular}} & $\ \ \ \ \ \ \ \   $ 1   $ \ \ \ \ \ \ \ \ \ $   | $\ $ \textcolor{red}{1899}  | \textcolor{blue}{1899}                               & \begin{tabular}{c} 0.289[6] \end{tabular}                                                & \begin{tabular}{c} -0.89[6] \end{tabular}                                              & 0.07[3]                                          & \begin{tabular}{c} 0.306[8] \end{tabular}                                            & \begin{tabular}{c} -1.03[5] \end{tabular}                                           & 0.06[2]                                       \\
                                                                                & $\ \ \ \ \ \ \ \ $ 1.3   $\ \ \  \ \ \  $   | $\ $ \textcolor{red}{1458}  | \textcolor{blue}{1458}                            & \begin{tabular}{c} 0.281[6]\end{tabular}                                                & \begin{tabular}{c} -0.77[7] \end{tabular}                                               & 0.06[3]                                       & \begin{tabular}{c} 0.288[5]\end{tabular}                                            & \begin{tabular}{c} -0.80[6] \end{tabular}                                           & 0.06[3]                                  \\

                                                                                & $\ \ \ \ \ \  $ 2   $ \ \ \  \ \ \ \ \ \  $   | $\ $ \textcolor{red}{235} $\ $  | \textcolor{blue}{235}                              & \begin{tabular}{c} 0.25[2] \end{tabular}                                                & \begin{tabular}{c} -0.45[5] \end{tabular}                                               & 0.06[4]                                      &  \begin{tabular}{c} 0.26[1] \end{tabular}                                            & \begin{tabular}{c} -0.43[9]\end{tabular}                                           & 0.05[3]                                          \\
                                                            \midrule

\multirow{6.22}{*}[3em]{\begin{tabular}[c]{@{}c@{}}Common  \\ w/ $R_{\rm{max}} \geqslant 2R_{\rm{e}}$ \\  \end{tabular}}   & $\ \ \ \ \ \  $ 1   $\ \  \ \ \ \ \ \  \  $   | $\ $ \textcolor{red}{235} $\ $  |  \textcolor{blue}{235}                              & \begin{tabular}{c} 0.31[2] \end{tabular}                                                & \begin{tabular}{c} -1.13[2] \\ \end{tabular}                                              & 0.08[2]                                       & \begin{tabular}{c} 0.30[1] \end{tabular}                                            & \begin{tabular}{c} -0.87[4] \end{tabular}                                           & 0.07[2]                                    \\
                                                                                & $\ \ \ \ \ \  $ 1.3   $\ \ \ \ \ \    $   | $\ $ \textcolor{red}{235} $\ $  |  \textcolor{blue}{235}                                                       & \begin{tabular}{c} 0.30[2]
                                                                                \end{tabular}                                                & \begin{tabular}{c} -0.90[9] \end{tabular}                                               & 0.07[4]                                     & \begin{tabular}{c} 0.27[1]\end{tabular}                                            & \begin{tabular}{c} -0.64[9] \end{tabular}                                           & 0.06[3]                                    \\
                                                              
\midrule

\multirow{6.24}{*}[3em]{\begin{tabular}[c]{@{}c@{}}\\ Rotators \end{tabular}}                                                     & $\ \ \ \ \ \  $ 1   $\ \ \ \ \ \ \ \  \  $   | $\ $ \textcolor{red}{879} $\ $  |  \textcolor{blue}{879}                                                         & {\cellcolor[rgb]{1,0.859,0.216}}\begin{tabular}{c} 0.285[7] \end{tabular}                & {\cellcolor[rgb]{1,0.859,0.216}}   \begin{tabular}{c} -0.82[7] \end{tabular}               & {\cellcolor[rgb]{1,0.859,0.216}}  0.05[2] & {\cellcolor[rgb]{0,0.898,0.62}} \begin{tabular}{c}0.284[6] \end{tabular}             & {\cellcolor[rgb]{0,0.898,0.62}}  \begin{tabular}{c} -0.77[6] \end{tabular}            & {\cellcolor[rgb]{0,0.898,0.62}}  0.05[2]          \\
                                                                                & $\ \ \ \ \ \  $ 1.3   $\ \ \ \ \ \   $   | $\ $ \textcolor{red}{644} $\ $  |  \textcolor{blue}{644}                                                       & {\cellcolor[rgb]{1,0.859,0.216}}\begin{tabular}{c} 0.285[7]\end{tabular}                & {\cellcolor[rgb]{1,0.859,0.216}} \begin{tabular}{c} -0.79[8] \end{tabular}               & {\cellcolor[rgb]{1,0.859,0.216}} 0.05[3]      & {\cellcolor[rgb]{0,0.898,0.62}}\begin{tabular}{c} \ 0.270[7]\end{tabular}             & {\cellcolor[rgb]{0,0.898,0.62}}\begin{tabular}{c} \ -0.60[7]\end{tabular}            & {\cellcolor[rgb]{0,0.898,0.62}}\ 0.05[3]    \\
                                                                                & $\ \ \ \ \ \  $ 2   $\ \ \ \    \ \  \ \ \  $   | $\ \  $ \textcolor{red}{79} $\  \ $  | $ \ $ \textcolor{blue}{79}                                                         & {\cellcolor[rgb]{1,0.859,0.216}} \begin{tabular}{c} 0.29[3] \end{tabular}                & {\cellcolor[rgb]{1,0.859,0.216}}\begin{tabular}{c} \ -0.82[9] \end{tabular}               & {\cellcolor[rgb]{1,0.859,0.216}}\ 0.04[3]        & {\cellcolor[rgb]{0,0.898,0.62}}\begin{tabular}{c} 0.28[2]\end{tabular}             & {\cellcolor[rgb]{0,0.898,0.62}}\begin{tabular}{c} \ -0.41[8] \end{tabular}            & {\cellcolor[rgb]{0,0.898,0.62}}\ 0.03[3]  

\\ 

\midrule 

\multirow{6.2}{*}[3em]{\begin{tabular}[c]{@{}c@{}}Rotators   \\ w/ $R_{\rm{max}} \geqslant 2R_{\rm{e}}$ \end{tabular}}                                                      & $\ \ \ \ \ \  $ 1   $\ \ \ \    \ \  \ \ \  $   | $\ \  $ \textcolor{red}{79} $\  \ $  | $\ $ \textcolor{blue}{79}                             & \begin{tabular}{c} 0.28[2]\end{tabular}                &    \begin{tabular}{c} -0.8[2] \end{tabular}               &   0.04[2] &  \begin{tabular}{c} 0.27[2]\end{tabular}             &   \begin{tabular}{c} -0.7[2] \end{tabular}            &   \ 0.04[2]      \\
                                                                                & $\ \ \ \ \ \  $ 1.3   $\ \ \ \    \ \      $   | $\ \  $ \textcolor{red}{79} $\  \ $  | $\ $ \textcolor{blue}{79}                           & \begin{tabular}{c} 0.31[2] \end{tabular}                &  \begin{tabular}{c} -1.1[3] \end{tabular}               & 0.03[3]           & \begin{tabular}{c} 0.26[2]\end{tabular}             & \begin{tabular}{c} -0.5[1]\end{tabular}            & \ 0.03[2]      \\

\midrule 
\midrule

\end{tabular*}
\end{table*}

\section{Tully-Fisher offset as a function of kinematic misalignment angle}
\label{sec:Appendix_A}

\begin{figure}
	\centering
	\includegraphics[width=\columnwidth]{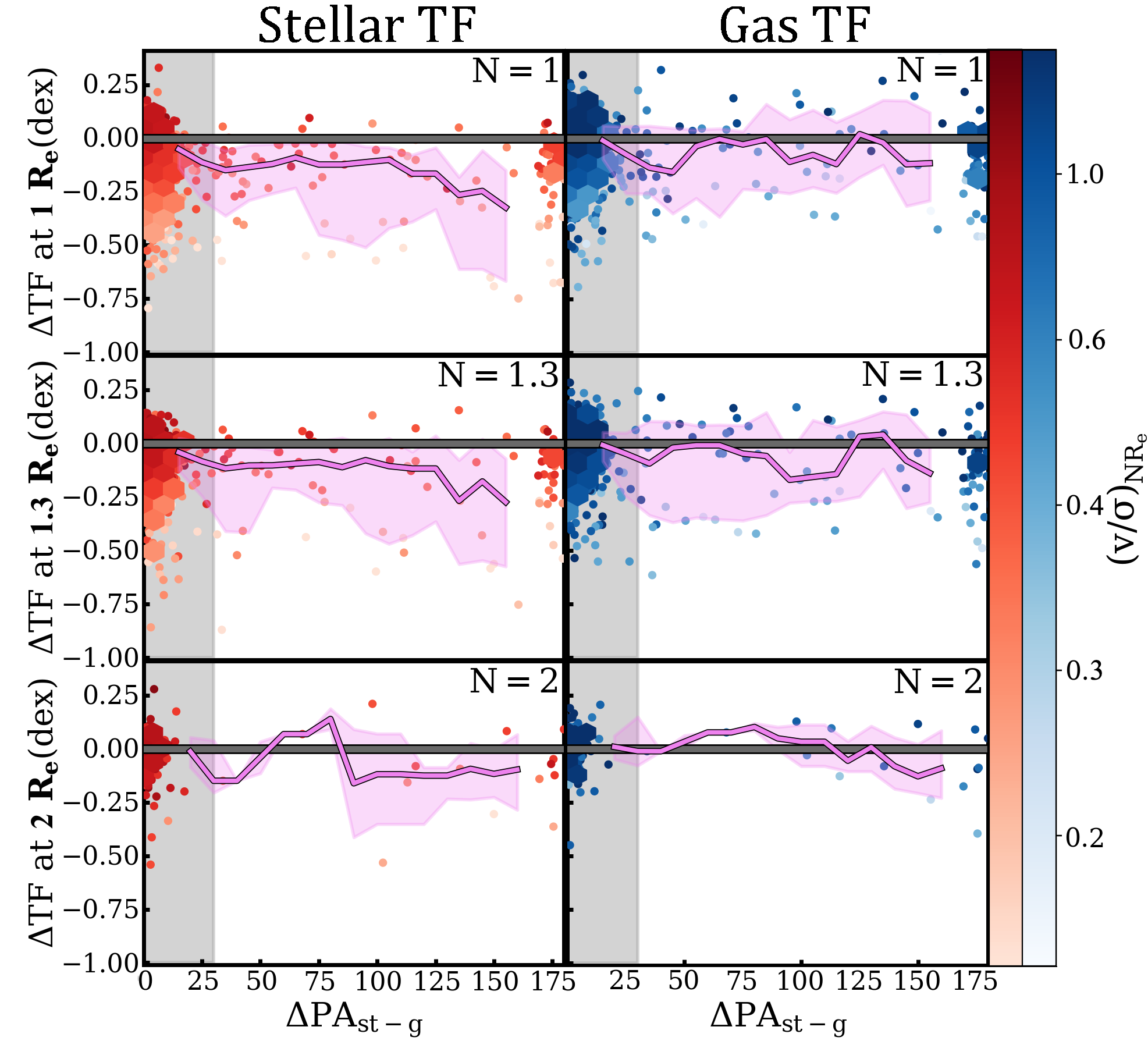}
    \caption{Dependence of \textcolor{red}{stellar} (\textbf{left}) and \textcolor{blue}{gas} (\textbf{right}) STFRs vertical offsets from a LSQ fit ($\Delta$TF) on the kinematic misalignment angle between stars and gas ($\rm{\Delta PA_{st-g}}$). Results are shown for the common kinematic sample reaching 1$R_{\rm{e}}$ (top row), 1.3$R_{\rm{e}}$ (middle row) and 2$R_{\rm{e}}$ (bottom row). Hexagonal bins are plotted for regions with at least 5 data points. The color coding shows the integrated $v/\sigma$ for stellar and gas kinematics within the respective radius, as indicated on the top right on each panel. The pink lines and shaded regions show the running medians and their scatter ($16^{\rm{th}}$ and $84^{\rm{th}}$ percentiles), while the grey shaded areas highlight the regions with $\rm{\Delta PA_{st-g}}<30^{\rm{o}}$, i.e. the galaxies which have aligned stellar and gas rotation \citep{ristea_sami_2022}. }
    \label{fig:dPA}
\end{figure}

In Fig. \ref{fig:dPA} we plot the offset from the stellar (left) and gas (right) STFR at 1$R_{\rm{e}}$ (top), 1.3$R_{\rm{e}}$ (middle) and 2$R_{\rm{e}}$ (bottom) as a function of the kinematic misalignment angle between stars and gas, $\rm{\Delta PA_{st-g}}$. There is no obvious linear trend between the plotted parameters, as reported in Fig. \ref{fig:grid_st} and Fig. \ref{fig:grid_g}. However, we find that 92 (65), 84 (67) and 73 (65) per cent of kinematically misaligned ($\rm{\Delta PA_{st-g}}\geqslant30^{\rm{o}}$, e.g. \citealt{ristea_sami_2022}) galaxies are found below the stellar (gas) STFRs ($\rm{\Delta TF < 0}$) at 1$R_{\rm{e}}$, 1.3 $R_{\rm{e}}$ and 2$R_{\rm{e}}$ respectively. We also note an association between the presence of misalignments and a decrease in integrated $v/\sigma$. The ratios between the median stellar and gas $v/\sigma$ for aligned and misaligned galaxies within 1$R_{\rm{e}}$, 1.3 $R_{\rm{e}}$ and 2$R_{\rm{e}}$ are $(2.1, 1.9, 2.0)_{\rm{stars}}$ and $(2.6, 2.7, 2.3)_{\rm{gas}}$, respectively.

These result for the gas STFR could be explained by a scenario in which misaligned gas accretion is increasing gas turbulence 
(\citealt{jimenez_physical_2023}), thus causing galaxies to scatter below the gas STFR. The same findings for the stellar kinematics could potentially be attributed to the fact that kinematic misalignments are more prevalent and longer-lived in early-type, high-dispersion galaxies (see e.g \citealt{bryant_sami_2019}; \citealt{ristea_sami_2022}). These galaxies are scattered below the stellar STFR due to an increase in dispersion compared to ordered rotation, potentially from a past misaligned accretion episode (from either a merger or the galaxy's outer halo) which has not yet stabilised.

\section*{Supporting information}
Additional supporting information can be found in the online version of this article:

\noindent \textbf{MaNGA kinematic catalogue}: Kinematic catalogue of MaNGA DR17 Galaxies, including measurements of stellar and gas rotational velocities at N$\times R_{\rm{e}}$ (N=1, 1.3, 2), and $v/\sigma$ ratios within the same radii.

%

\bsp	
\label{lastpage}
\end{document}